\documentclass[aps,prl,twocolumn,floatfix,showpacs,superscriptaddress]{revtex4-2}
\usepackage{graphics}
\usepackage{epsfig}
\usepackage{times}
\usepackage{bm}
\usepackage{braket}
\usepackage{color}

\usepackage{amsmath}	
\begin{document}
\title{One-dimensional projection of two-dimensional systems using spiral boundary conditions}
\author{Masahiro Kadosawa}
\affiliation{Department of Physics, Chiba University, Chiba 263-8522, Japan}
\author{Masaaki Nakamura}
\affiliation{Department of Physics, Ehime University, Ehime 790-8577, Japan}
\author{Yukinori Ohta}
\affiliation{Department of Physics, Chiba University, Chiba 263-8522, Japan}
\author{Satoshi Nishimoto}
\affiliation{Department of Physics, Technical University Dresden, 01069 Dresden, Germany}
\affiliation{Institute for Theoretical Solid State Physics, IFW Dresden, 01069 Dresden, Germany}

\date{\today}

\begin{abstract}
We introduce spiral boundary conditions (SBCs) as a useful tool for
handling the shape of finite-size periodic clusters.
Using SBCs, a lattice model for more than two dimensions can be exactly
projected onto a one-dimensional (1D) periodic chain with translational invariance. Hence, the existing 1D techniques such as density-matrix
renormalization group (DMRG), bosonization, Jordan–Wigner transformation,
etc., can be effectively applied to the projected 1D model. First, we describe
the 1D projection scheme for the two-dimensional (2D) square- and
honeycomb-lattice tight-binding models in real and momentum space.
Next, we discuss how the density of states and the ground-state energy
approach their thermodynamic limits. Finally, to demonstrate the utility
of SBCs in DMRG simulations, we estimate the magnitude of staggered magnetization
of the 2D XXZ Heisenberg model as a function of XXZ anisotropy.

\end{abstract}

\maketitle

{\it Introduction.}
In condensed matter physics, theoretical research is usually carried
out based on the statistical mechanical formulation of either lattice or
continuum models which describe the microscopic structure of solids.
In general, a lattice model is a cluster of lattice points corresponding
to the positions of aligned atoms in a crystal~\cite{Ashcroft1976}.
The Hamiltonian is typically expressed as a countable set of lattice points
or bonds, because finite degrees of freedom such as spin, charge, hole, etc
are assigned in each lattice point. Therefore, unlike in the continuum limit
with huge degrees of freedom, a lattice model is rather suitable for computer
simulations. Typical examples of lattice model are the Hubbard
model~\cite{Hubbard1963}, the Heisenberg model~\cite{Heisenberg1928},
the Kondo-lattice model~\cite{Kondo1964}, and the Kitaev
model~\cite{Kitaev2006}. A microscopic starting point to understand
the electronic and/or magnetic properties of solids is provided by solving
those kinds of lattice models analytically or numerically.

When studying such a model in numerical simulations, we usually put it
on a lattice of finite size. Then, an extrapolation of the result to an
infinite system is considered if necessary. However, since the total
degrees of freedom of the system increases exponentially with lattice
size, the geometry of the cluster can be strongly restricted especially for
systems in more than two dimensions. In such cases, the management
of boundary conditions is crucial to ensure "correct" simulations.
Either periodic boundary conditions (PBCs), open boundary conditions (OBCs),
or a combination of them, such as e.g., a cylinder, are typically used.
Nevertheless, a naive choice of boundary conditions could easily give
rise to a situation where the lowest-energy state with a small cluster
is not relevant to the ground state (GS) in the thermodynamic limit, instead
of systematic errors due to the finite-size effects. This issue could be
addressed for example by the sorting of states with, e.g., quantum numbers,
momentum, and parity, as explicitly done in level
spectroscopy~\cite{Okamoto1992}, or by controlling the open edges for a particular
state~\cite{S1kagome}. However, these approaches are not always successful.

A simple alternative way to resolve or reduce the above issue is by the use
of spiral boundary conditions (SBCs). As explained below, SBCs enable us to 
represent two-dimensional (2D) lattice sites by a one-dimensional (1D) array.
This method was originally used to optimize the computational cost in Monte Carlo simulations~\cite{Newman1999} but it also allows us to efficiently apply existing 1D techniques such as the density-matrix renormalization group (DMRG)~\cite{White1992}, bosonization~\cite{Gogolin2004}, and Jordan–Wigner
transformation~\cite{Chen2008}, etc. SBCs have been also introduced for
extending the Lieb-Schultz-Mattis theorem to higher dimensions~\cite{Yao2020}
and for discussing the GS degeneracy in the thermodynamic
limit~\cite{Yamada2022}. In this Letter, we thus propose SBCs as a useful
tool for handling the shape of finite-size periodic clusters. As practical
examples, we describe the 1D projections of the 2D square- and
honeycomb-lattice tight-binding (TB) models in real and momentum space.
Then, we present how the density of states (DOS) and the GS energy approach
their thermodynamic limits. Furthermore, in order to demonstrate the
utility of SBCs, we calculate the GS energy of the 2D half-filled Hubbard
model and the magnitude of staggered magnetization of the 2D XXZ Heisenberg
model using the DMRG method.  Their estimations had been longstanding problems
and were only recently settled~\cite{LeBlanc2015,Sandvik2010}.
In our DMRG calculations, we keep up to $12\,000$ density-matrix
eigenstates and the typical discarded weight is smaller than $\sim 10^{-5}$.
More detailed data are given in the Supplemental Material.

\begin{figure}[tbh]
	\centering
	\includegraphics[width=1.0\linewidth]{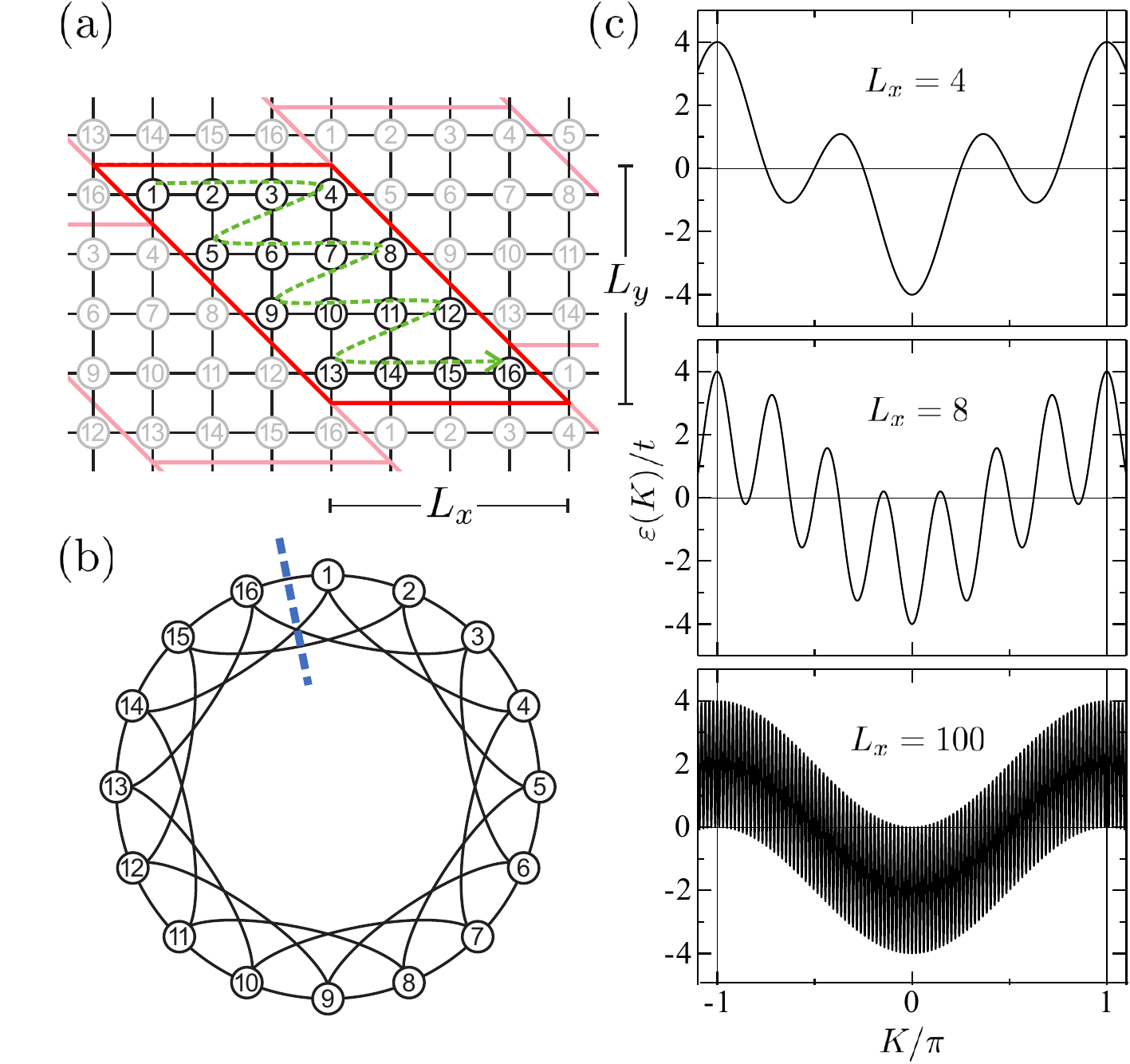}
	\caption{
		(a) 2D square-lattice cluster with $4 \times 4$ sites, where
		the region framed by the red line is the original cluster.
		(b) 1D representation of the cluster (a) by numbering sites
		along the green line. This is a periodic chain, and an open chain is
		created by cutting $L_x$ bonds between two sites (dotted line).
		(c) $L_x$ dependence of the energy dispersion $\varepsilon(K)$,
		where the limit $L_y\to\infty$ is taken.
	}
	\label{lattice_sq}
\end{figure}

{\it Projection of 2D cluster onto 1D chain using SBCs.}
SBCs are a variation of the idea of PBCs. They provide a way
to map lattice models for more than two dimensions onto 1D periodic chains
with translation symmetry. As an illustration, let us consider
the TB model on a square lattice with $L_x \times L_y$ sites.
When a PBC is applied to its finite-size system, we usually use a rectangular
cluster and sites on one edge of the cluster are assumed to be neighbors
of the corresponding sites on the opposite edge. However, this choice may
be arbitrarily deformed as long as the boundaries have the correct topology
of state. Thus, we now make a particular choice of boundaries shown
in Fig.~\ref{lattice_sq}(a), where all sites are traced along the dashed line
in a spiral manner. In this way, the original 2D cluster can be exactly
reproduced as a 1D chain with nearest- and $(L_x-1)$th-neighbor hopping integrals [Fig.~\ref{lattice_sq}(b)], where the translational symmetry is
preserved. The Hamiltonian is written as 
${\cal H}_{\rm sq, 0} =-t \sum_\sigma\sum_{i=1}^{L_xL_y}(c^\dagger_{i,\sigma}c_{i+1,\sigma}+c^\dagger_{i,\sigma}c_{i+(L_x-1),\sigma}+{\rm H.c.})$, where $c_{i,\sigma}$ is
an annihilation operator of an electron with spin $\sigma$ at site $i$, and
$t$ is the nearest-neighbor hopping integral in the original 2D model.
Its Fourier transform
leads to ${\cal H}_{{\rm sq},K} =-2t \sum_{K,\sigma}[\cos K + \cos (L_x-1)K] c^\dagger_{K,\sigma}c_{K,\sigma}$, where $K$ is the momentum defined along
the projected 1D periodic chain and
$c_{K,\sigma}=(1/\sqrt{L_xL_y})\sum_i \exp(iKr_i)c_{i,\sigma}$.
Since the sites are ordered along a "snakelike" path shown in Fig.~\ref{lattice_sq}(a),
the original 2D momenta $(k_x,k_y)=\left(\frac{2\pi}{L_x}n_x,\frac{2\pi}{L_y}n_y\right)$
($n_x=0,1,\dots,L_x-1; n_y=0,1,\dots,L_y-1$) are transferred to the 1D
momentum $K=\frac{2\pi}{L_xL_y}n$ with $n=n_x+L_xn_y=0,1,\dots,L_xL_y-1$.

The $L_x$ dependence of the energy dispersion
$\varepsilon(K)=-2t[\cos K + \cos (L_x-1)K]$ is shown
in Fig.~\ref{lattice_sq}(c), where the limit $L_y\to\infty$ is taken
to obtain a continuous dispersion with $K$. Reflecting the snakelike
order of sites in the original 2D cluster, the dispersion is oscillating
as a function of $K$. In the large $L_x$ limit it turns out to be a beltlike
dispersion relation which is interpreted as the projected band structure
of the square-lattice TB model onto a Cartesian axis, i.e., $x$ or $y$.
The original 2D Fermi surface is represented as a "Fermi line."
For example, when the Fermi level is set at $0<\varepsilon_{\rm F}<4$,
two separate unoccupied regions correspond to the hole pockets.

Let us see the case of half filling. The dispersion is particle-hole
symmetric and there are $2L_x-2$ Fermi points. The GS energy $E_0$ can
be calculated by carrying out the single-particle energy summation over
the $L_x-1$ regions with $\varepsilon(K)<0$,
\begin{align}
\frac{E_0}{L_xL_y}&=-\frac{4}{\pi}\int_{\varepsilon(K)<0}(\cos K+\cos[(L_x-1)K])dK\\
&=-\frac{4}{\pi}\sum^{L_x/2}_{j=1}\left[\sin K+\frac{\sin[(L_x-1)K]}{L_x-1}\right]^b_a\\
\nonumber
&\underset{L_x,L_y\to\infty}{\longrightarrow}-\frac{16}{\pi^2},
	\label{local_chi}
\end{align}
where $a=\min(0,\frac{2j-3}{L_x-2}\pi)$ and $b=\frac{2j-1}{L_x}\pi$.
This energy coincides with that of the infinite-size system. Therefore,
we can confirm that the finite-size systems under SBCs are adiabatically
connected to the thermodynamic limit. More details are given in the
Supplemental Material.

\begin{figure}[tbh]
	\centering
	\includegraphics[width=1.0\linewidth]{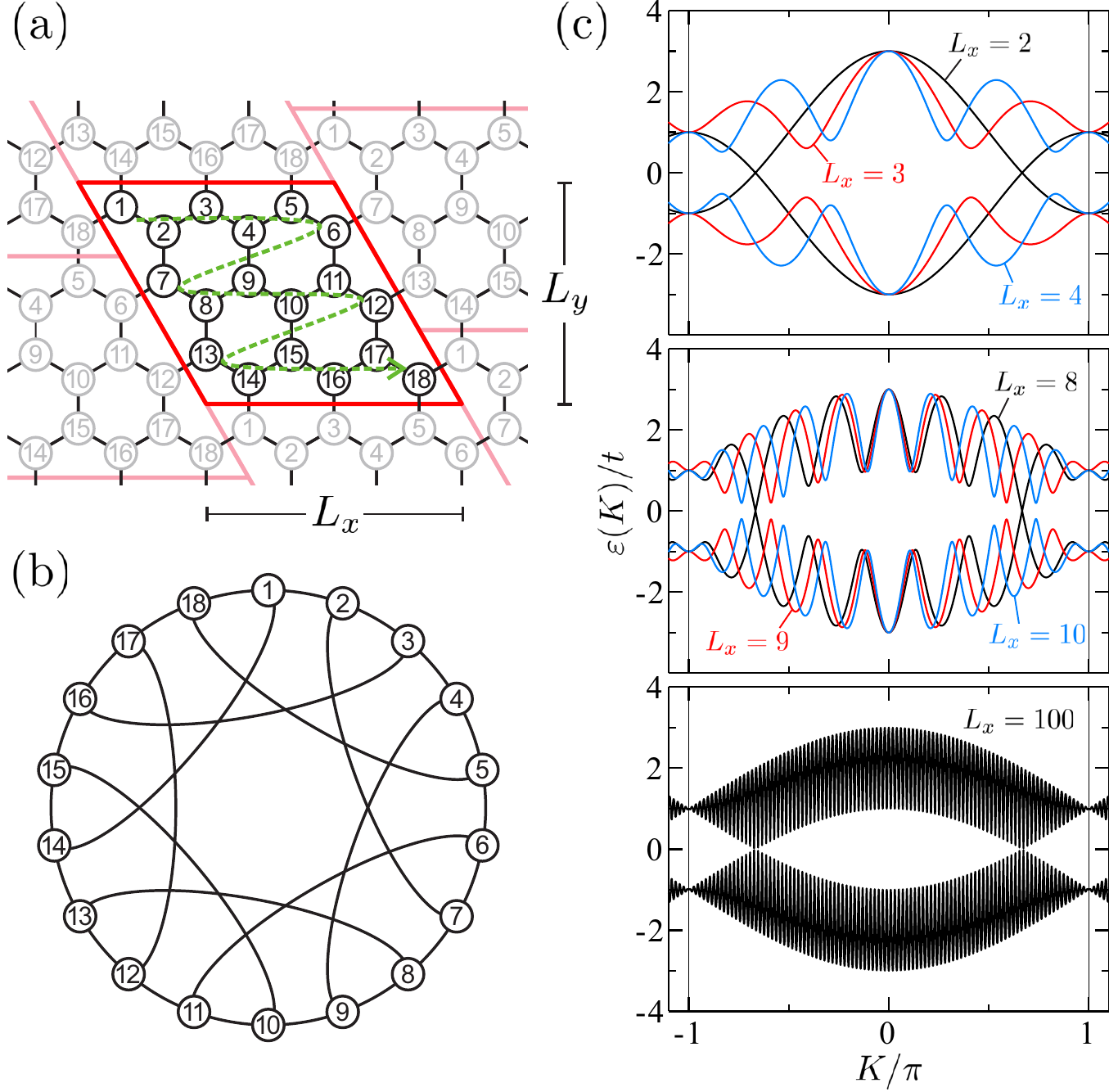}
	\caption{
		(a) 2D honeycomb-lattice cluster with $3 \times 3$ unit cells,
		where a region framed by the red line is the original cluster.
		(b) 1D representation of the cluster (a) by numbering sites along
		the green line.
		(c) $L_x$-dependence of the energy dispersion $\varepsilon(K)$,
		where the limit $L_y\to\infty$ is taken. The upper and lower bands
		are degenerate at $K=\pm\frac{2}{3}\pi$ when $2L_x-1=3M$ ($M$: integer).
	}
	\label{lattice_hon}
\end{figure}

Note that the way of 1D projection using SBCs is not unique. This means
that the modulation of the wave function can be controlled more flexibly 
than PBCs~\cite{Nakamura2021}. In other words, a periodic
system consistent with an arbitrary commensurate ordering vector can be 
created by tuning the shape of the finite-size cluster and its alignment.
Similar boundary conditions have been used in some
numerical calculations to manage a limited periodicity of small
clusters~\cite{Nishiyama2009,Miyata2021}. Other examples of SBC usage are
given in the Supplemental Material.

{\it Honeycomb-lattice TB model under SBCs.}
Another interesting example is the honeycomb-lattice TB model. The choice of
spiral boundaries and the corresponding projected 1D chain are shown in
Figs.~\ref{lattice_hon}(a) and \ref{lattice_hon}(b), respectively. The Hamiltonian of the
projected 1D chain is written as
${\cal H}_{\rm hon,0} =-t \sum_{i=1}^{L_xL_y}\sum_\sigma(c^\dagger_{2i-1,\sigma}c_{2i,\sigma}+c^\dagger_{2i,\sigma}c_{2i+1,\sigma}+c^\dagger_{2i,\sigma}c_{2i+(2L_x-1),\sigma}+{\rm H.c.})$.
For the limit of $L_y\to\infty$ the energy dispersion is written as
$\varepsilon(K)=\pm t \sqrt{1+4\cos^2\frac{K}{2}+4\cos\frac{K}{2}\cos\frac{(2L_x-1)K}{2}}$. The $L_x$ dependence of $\varepsilon(K)$ is shown in
Fig.~\ref{lattice_hon}(c). The upper and lower bands are degenerate at
$K=\pm\frac{2}{3}\pi$ only when $2L_x-1=3M$, where $M$ is an integer.
In the limit of $L_x,L_y\to\infty$, the band structure is equivalent to the
projected one of 2D graphene onto a zigzag axis. Accordingly, the Dirac
points of the original 2D graphene are reproduced at $K=\pm\frac{2}{3}\pi$.

\begin{figure}[tbh]
\centering
\includegraphics[width=1.0\linewidth]{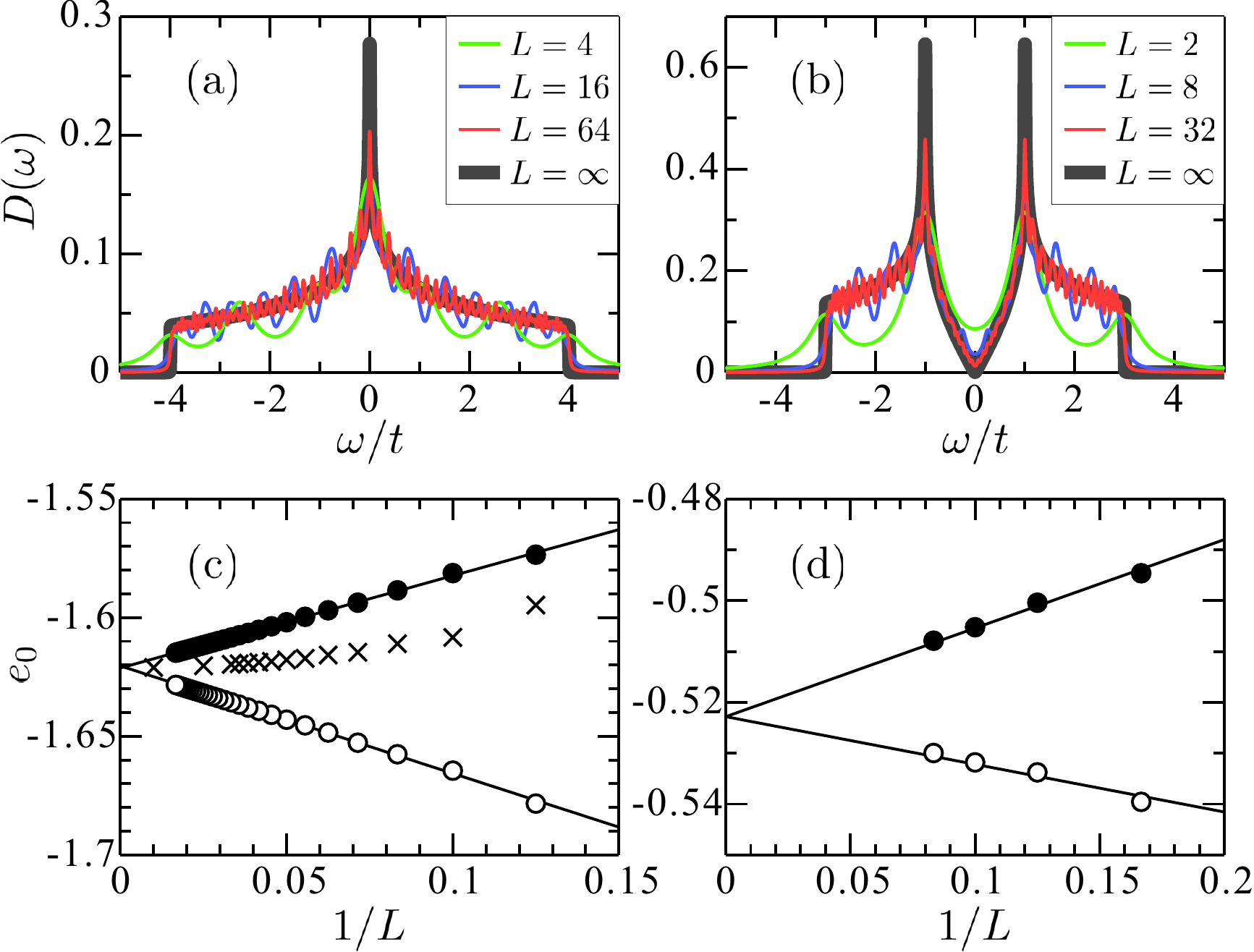}
\caption{
$L$ dependence of the density of states for (a) square- and (b)
honeycomb-lattice TB models, where a broadening of the $\delta$ peak
$1.6/L$ and $0.8/L$ is introduced, respectively.
Finite-size scaling analysis of the ground-state energy for the
half-filled square-lattice Hubbard model at (c) $U=0$ and (d) $U/t=8$.
Solid and open circles denote the energy per two bonds and per site
calculated with open chains, respectively. Crosses denote the energy per
site calculated with periodic chains.
}
\label{scaling}
\end{figure}

{\it Approach to the thermodynamic limit.}
It is informative to see how the DOS and the GS energy approach their
thermodynamic limits. For simplicity, hereafter we consider the case of $L_x=L_y=L$.
The evolution of the DOS with $L$ for the square- and
honeycomb-lattice TB models is shown in Figs.~\ref{scaling}(a) and \ref{scaling}(b),
respectively. With increasing $L$, they are smoothly connected to the
thermodynamic limit ones.
Also, the overall shape including the van Hove singularity can be
approximately reproduced even with a relatively small cluster. It is
because the degeneracy of the energy levels in a finite-size PBC cluster
is lifted due to the partial breaking of its rotation symmetry by SBCs.
For a square-lattice cluster with $L \times L$ sites, the number of
independent momenta is $\frac{L^2}{2}+1$ under SBCs and
$\frac{L^2}{8}+\frac{3L}{4}+1$ under PBCs. More details are discussed in the Supplemental Material.

In Fig.~\ref{scaling}(c) a finite-size scaling analysis of the GS energy 
for the half-filled square-lattice TB model with the projected 1D periodic
chains is shown. The data points are analytically obtained. As expected, the energy
per site ($e_0$) quadratically approaches $-\frac{16}{\pi^2}$ as a function
of $\frac{1}{L}$. It is also interesting to see the scaling behavior when
open chains are used. An open chain is created by cutting $L$ bonds between
two neighboring sites of the periodic chain [see Fig.~\ref{lattice_sq}(b)]. 
Note though that the number of missing bonds is reduced from $2L$ in the
original 2D PBC cluster to $L$. Nevertheless, since the ratio of the number
of bonds per site deviates from $2$ due to the missing bonds for finite-size
open chains, it is convenient to estimate the GS energy in two different ways:
One is the energy per site and the other is that per two bonds. As shown
in Fig.~\ref{scaling}(c), both of them are extrapolated almost linearly
to the thermodynamic limit. One of them is extrapolated from the higher-energy
side with decreasing $\frac{1}{L}$ and the other from the lower-energy side,
so that this makes the scaling analysis more reliable. Eventually,
the scaling behavior with open chains seems to be even more simple than
that with periodic chains.

\begin{figure}[tbh]
	\centering
	\includegraphics[width=1.0\linewidth]{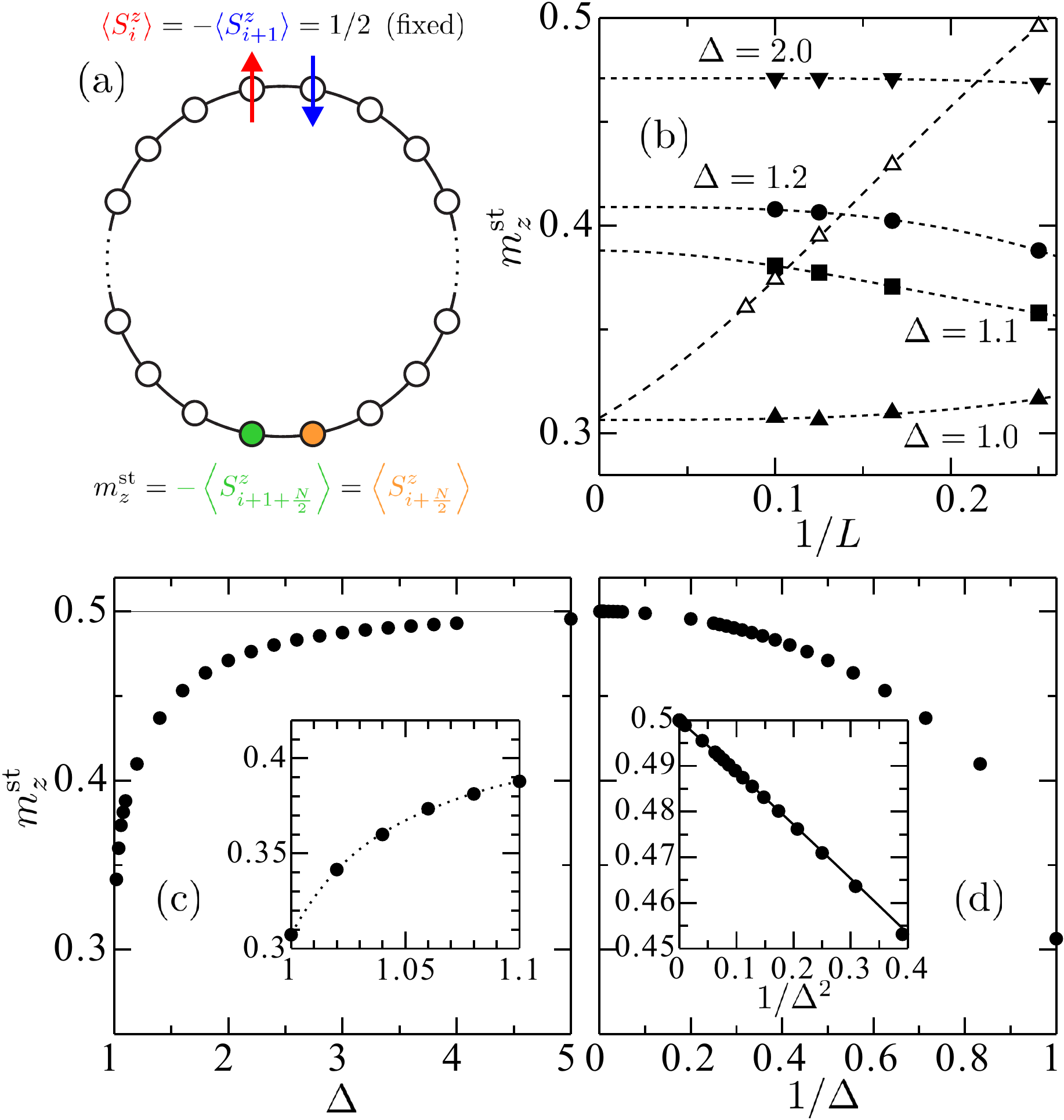}
	\caption{
		(a) Schematic picture of the 1D periodic chain used for the DMRG
		calculations of staggered magnetization $m_z^{\rm st}$.
		(b) Finite-size scaling analysis of $m_z^{\rm st}$. Solid symbols
		denote the data points estimated with spin rotation symmetry
		breaking, and open triangles denote those estimated from the
		static structure factor (see text). (c), (d) Extrapolated values
		of $m_z^{\rm st}$ in the thermodynamic limit. Insets: Enlarged
		views near $\Delta=1$ and $1/\Delta=0$, respectively. The solid line
		in the inset of (d) shows a fitting by a polynomial function
		with $1/\Delta$.
	}
	\label{fig:mstXXZ}
\end{figure}

{\it Application of SBCs in DMRG calculations.}
In DMRG simulations for a 2D system, it is not easy to obtain
physical quantities in the thermodynamic limit because not only
are their implementations challenging even with finite-size clusters, but also
the finite-size scaling analysis must be performed along two orientations,
e.g., the $x$ and $y$ directions. This issue can be somewhat alleviated
by applying the above 1D projection scheme. In order to demonstrate this,
we here present two examples of DMRG simulations for a 2D system.

The first example is the GS energy of the 2D half-filled
Hubbard model on a square lattice, whose Hamiltonian is
${\cal H}={\cal H}_{\rm sq,0}+U\sum_i n_{i,\uparrow}n_{i,\downarrow}$, where $n_{i,\sigma}=c^\dagger_{i,\sigma}c_{i,\sigma}$. In Fig.~\ref{scaling}(d)
the finite-size scaling of GS energy for $U=8$ is performed.
As is the case in the TB model, open chains are used, so that
the extrapolation to the thermodynamic limit seems to be straightforward.
It leads to $e_0=-0.5228$ by linear fitting. This energy is only slightly
higher than $e_0=-0.5241$ estimated by DMRG calculations with
infinite-length cylinders~\cite{LeBlanc2015}. Perhaps the extrapolation
in the circumferential direction may contain some uncertainty due to the
unsettled scaling function with several data points.

The second example is the spontaneous staggered magnetization of the 2D
XXZ Heisenberg model on a square lattice, whose Hamiltonian is
${\cal H}=\sum_{\langle i,j \rangle} (S^x_iS^x_j+S^y_iS^y_j+\Delta S^z_iS^z_j)$,
where $S^\gamma_i$ are the spin-$\frac{1}{2}$ operators associated with 
site $i$, $\Delta$ is the anisotropy parameter, and the sum
$\langle i,j \rangle$ runs over all nearest-neighbor pairs. We here use
periodic chains. As sketched in Fig.~\ref{fig:mstXXZ}(a), the $z$ components
of spins at sites $i$ and $i+1$ are fixed to $\frac{1}{2}$ and $-\frac{1}{2}$,
respectively, and the spin moments at the farthest two sites from the fixed
spins are measured:
$m_z^{\rm st}=-\langle S^z_{i+1+\frac{N}{2}}\rangle=\langle S^z_{i+\frac{N}{2}}\rangle$. Several examples of the finite-size scaling analysis
are shown in Fig.~\ref{fig:mstXXZ}(b), where the spin moments are calculated
using periodic chains with lengths up to $N=L^2=100$ sites. For the isotropic
case ($\Delta=1$), we obtain $m_z^{\rm st}=0.3071 \pm 0.0005$ in the
thermodynamic limit. This value is reasonably close to the previous DMRG 
($m_z^{\rm st}=0.3067$)~\cite{White2007} and quantum Monte Carlo
($m_z^{\rm st}=0.307\,43$)~\cite{Sandvik2010} estimations. The magnetization
increases with increasing $\Delta$. The extrapolated values of $m_z^{\rm st}$
are plotted as a function of $\Delta$ in Fig.~\ref{fig:mstXXZ}(c).
The overall behavior is basically consistent with the previous studies
~\cite{Weihong1991-1,Hamer1992,Lin2001,Bishop2017}. However, a singularity
near $\Delta=1$, $m_z^{\rm st}=\sum_{n=0}^{\infty}\mu_n(1-\Delta^{-2})^{n/2}$,
predicted by the spin-wave theory~\cite{Weihong1991-1,Weihong1991-2,Huse1988}
is not confirmed in our results. This is consistent with results from
the coupled cluster method~\cite{Bishop2017}. A more detailed
analysis is given in Ref.~\cite{Kadosawa2023}. On the other hand, as shown in Fig.~\ref{fig:mstXXZ}(d), our data in the large-$\Delta$ region
($0\le1/\Delta\le0.05$) can be fitted by
$2m_z^{\rm st}=1+m_2/\Delta^2+m_4/\Delta^4+m_6/\Delta^6$ with
$m_2=-0.222\,222\,225$, $m_4=-0.035\,554\,273\,6$, and
$m_6=-0.018\,966\,381\,0$. This agrees well with the series expansions
$2m_z^{\rm st}=1-(2/9)/\Delta^2-(8/225)/\Delta^4-0.018\,942\,58/\Delta^6+
{\cal O}(1/\Delta^8)\qquad$($2/9=0.222\,222\,22\dots$, $8/225=0.035\,555\,5\dots$)
~\cite{Weihong1991-1}.

Finally, we present another option to calculate the staggered magnetization.
In the above estimations, the spin rotation symmetry is broken by design.
Although it makes the DMRG calculations more stable, an equally precise
estimation of $m_z^{\rm st}$ is also possible without such explicit
symmetry breaking. Using open chains, we can accurately estimate the
staggered magnetization from the static structure factor
$(m_z^{\rm st})^2=\lim_{L\to\infty}(1/L^2)\sum_{ij} (-1)^{i-j} \langle \bm{S}_i\cdot\bm{S}_j \rangle$, where the sum is taken over the open chain.
The finite-size scaling analysis using the chains with lengths up to
$12\times12$ is given in Fig.~\ref{fig:mstXXZ}(b).
We obtain $m_z^{\rm st}=0.307\,62 \pm 0.0032$ in the thermodynamic limit.

{\it Summary.}
Applying SBCs, lattice models for more than two dimensions can be exactly
projected onto 1D periodic chains with translational invariance. 
In the projected 1D chain, each lattice site is indexed by a single
coordinate instead of two coordinates in the original 2D PBC cluster,
so that we only have to perform a finite-size scaling analysis along the
chain direction to obtain a physical quantity in the thermodynamic limit.
As practical examples, we first explained how the 2D square- and
honeycomb-lattice TB models are expressed as 1D periodic systems in both
real and momentum space. Then, the evolution of the DOS with increasing
cluster size as well as a finite-size scaling analysis of the GS energy
to the thermodynamic limit was shown. Finally, in order to demonstrate the utility of this 1D projection scheme in DMRG simulations, we calculated
the magnitude of staggered magnetization in the 2D XXZ Heisenberg model
on a square lattice.

The 1D projection scheme using SBCs can be extended to further research. 
Since the projected 1D chain has translational symmetry,
all of the so-called (local) $A$ tensors are set to be equivalent in a matrix
product state. As a result, quantum entanglement is uniformly distributed
over the projected 1D chain. It is also important that the distance of the
longest bonds is minimized. These conditions enable us to optimally perform
DMRG calculations, and also allow us to use the existing techniques such as
infinite DMRG and transfer-matrix renormalization group.
Though only two kinds of 2D lattices are considered in this Letter, a similar
1D projection is possible for any periodic lattices in more than two dimensions~\cite{Newman1999}.
Moreover, in most cases SBCs are expected to practically give an easier finite-size scaling analysis than the cylinder and PBCs.
To clarify the advantages of SBCs in DMRG simulations, the comparison of
performance with other boundary conditions is discussed in the Supplemental
Material.

{\it Acknowledgements.}
We thank Ulrike Nitzsche for technical support. This work was supported by Grants-in-Aid for Scientific Research from JSPS (Projects No. JP20H01849, No. JP20K03769, and No. JP21J20604).
M.K. acknowledges support from the JSPS Research Fellowship for Young Scientists.
M.N. acknowledges the Visiting Researcher’s Program of the Institute
for Solid State Physics, The University of Tokyo, and the research fellow
position of the Institute of Industrial Science, The University of Tokyo.
S.N. acknowledges support from SFB 1143 project A05
(project-id 247310070) of the Deutsche Forschungsgemeinschaft.

\bibliography{SBC}

\begin{thebibliography}{26}%
\makeatletter
\providecommand \@ifxundefined [1]{%
 \@ifx{#1\undefined}
}%
\providecommand \@ifnum [1]{%
 \ifnum #1\expandafter \@firstoftwo
 \else \expandafter \@secondoftwo
 \fi
}%
\providecommand \@ifx [1]{%
 \ifx #1\expandafter \@firstoftwo
 \else \expandafter \@secondoftwo
 \fi
}%
\providecommand \natexlab [1]{#1}%
\providecommand \enquote  [1]{``#1''}%
\providecommand \bibnamefont  [1]{#1}%
\providecommand \bibfnamefont [1]{#1}%
\providecommand \citenamefont [1]{#1}%
\providecommand \href@noop [0]{\@secondoftwo}%
\providecommand \href [0]{\begingroup \@sanitize@url \@href}%
\providecommand \@href[1]{\@@startlink{#1}\@@href}%
\providecommand \@@href[1]{\endgroup#1\@@endlink}%
\providecommand \@sanitize@url [0]{\catcode `\\12\catcode `\$12\catcode
  `\&12\catcode `\#12\catcode `\^12\catcode `\_12\catcode `\%12\relax}%
\providecommand \@@startlink[1]{}%
\providecommand \@@endlink[0]{}%
\providecommand \url  [0]{\begingroup\@sanitize@url \@url }%
\providecommand \@url [1]{\endgroup\@href {#1}{\urlprefix }}%
\providecommand \urlprefix  [0]{URL }%
\providecommand \Eprint [0]{\href }%
\providecommand \doibase [0]{https://doi.org/}%
\providecommand \selectlanguage [0]{\@gobble}%
\providecommand \bibinfo  [0]{\@secondoftwo}%
\providecommand \bibfield  [0]{\@secondoftwo}%
\providecommand \translation [1]{[#1]}%
\providecommand \BibitemOpen [0]{}%
\providecommand \bibitemStop [0]{}%
\providecommand \bibitemNoStop [0]{.\EOS\space}%
\providecommand \EOS [0]{\spacefactor3000\relax}%
\providecommand \BibitemShut  [1]{\csname bibitem#1\endcsname}%
\let\auto@bib@innerbib\@empty
\bibitem [{\citenamefont {Ashcroft}\ and\ \citenamefont
  {Mermin}(1976)}]{Ashcroft1976}%
  \BibitemOpen
  \bibfield  {author} {\bibinfo {author} {\bibfnamefont {N.~W.}\ \bibnamefont
  {Ashcroft}}\ and\ \bibinfo {author} {\bibfnamefont {N.~D.}\ \bibnamefont
  {Mermin}},\ }\href@noop {} {\emph {\bibinfo {title} {Solid state physics}}}\
  (\bibinfo  {publisher} {Saunders College Publishing, Philadelphia},\ \bibinfo
  {year} {1976})\BibitemShut {NoStop}%
\bibitem [{\citenamefont {Hubbard}(1963)}]{Hubbard1963}%
  \BibitemOpen
  \bibfield  {author} {\bibinfo {author} {\bibfnamefont {J.}~\bibnamefont
  {Hubbard}},\ }\bibfield  {title} {\bibinfo {title} {Electron correlations in
  narrow energy bands},\ }\href {https://doi.org/10.1098/rspa.1963.0204}
  {\bibfield  {journal} {\bibinfo  {journal} {Proc. R. Soc. London, Ser. A}\
  }\textbf {\bibinfo {volume} {276}},\ \bibinfo {pages} {238} (\bibinfo {year}
  {1963})}\BibitemShut {NoStop}%
\bibitem [{\citenamefont {Heisenberg}(1928)}]{Heisenberg1928}%
  \BibitemOpen
  \bibfield  {author} {\bibinfo {author} {\bibfnamefont {W.~J.}\ \bibnamefont
  {Heisenberg}},\ }\bibfield  {title} {\bibinfo {title} {Zur theorie des
  ferromagnetismus},\ }\href {https://doi.org/10.1007/BF01328601} {\bibfield
  {journal} {\bibinfo  {journal} {Zeitschrift f\"ur Physik}\ }\textbf {\bibinfo
  {volume} {49}},\ \bibinfo {pages} {619} (\bibinfo {year} {1928})}\BibitemShut
  {NoStop}%
\bibitem [{\citenamefont {Kondo}(1964)}]{Kondo1964}%
  \BibitemOpen
  \bibfield  {author} {\bibinfo {author} {\bibfnamefont {J.}~\bibnamefont
  {Kondo}},\ }\bibfield  {title} {\bibinfo {title} {Resistance minimum in
  dilute magnetic alloys},\ }\href {https://doi.org/10.1143/PTP.32.37}
  {\bibfield  {journal} {\bibinfo  {journal} {Prog. Theor. Phys.}\ }\textbf
  {\bibinfo {volume} {32}},\ \bibinfo {pages} {37} (\bibinfo {year}
  {1964})}\BibitemShut {NoStop}%
\bibitem [{\citenamefont {Kitaev}(2006)}]{Kitaev2006}%
  \BibitemOpen
  \bibfield  {author} {\bibinfo {author} {\bibfnamefont {A.}~\bibnamefont
  {Kitaev}},\ }\bibfield  {title} {\bibinfo {title} {Anyons in an exactly
  solved model and beyond},\ }\href
  {https://doi.org/https://doi.org/10.1016/j.aop.2005.10.005} {\bibfield
  {journal} {\bibinfo  {journal} {Ann. Phys.}\ }\textbf {\bibinfo {volume}
  {321}},\ \bibinfo {pages} {2} (\bibinfo {year} {2006})}\BibitemShut {NoStop}%
\bibitem [{\citenamefont {Okamoto}\ and\ \citenamefont
  {Nomura}(1992)}]{Okamoto1992}%
  \BibitemOpen
  \bibfield  {author} {\bibinfo {author} {\bibfnamefont {K.}~\bibnamefont
  {Okamoto}}\ and\ \bibinfo {author} {\bibfnamefont {K.}~\bibnamefont
  {Nomura}},\ }\bibfield  {title} {\bibinfo {title} {Fluid-dimer critical point
  in {$S=1/2$} antiferromagnetic {H}eisenberg chain with next nearest neighbor
  interactions},\ }\href
  {https://doi.org/https://doi.org/10.1016/0375-9601(92)90823-5} {\bibfield
  {journal} {\bibinfo  {journal} {Phys. Lett. A}\ }\textbf {\bibinfo {volume}
  {169}},\ \bibinfo {pages} {433} (\bibinfo {year} {1992})}\BibitemShut
  {NoStop}%
\bibitem [{\citenamefont {Nishimoto}\ and\ \citenamefont
  {Nakamura}(2015)}]{S1kagome}%
  \BibitemOpen
  \bibfield  {author} {\bibinfo {author} {\bibfnamefont {S.}~\bibnamefont
  {Nishimoto}}\ and\ \bibinfo {author} {\bibfnamefont {M.}~\bibnamefont
  {Nakamura}},\ }\bibfield  {title} {\bibinfo {title} {Non-symmetry-breaking
  ground state of the {$S=1$} {Heisenberg} model on the kagome lattice},\
  }\href {https://doi.org/10.1103/PhysRevB.92.140412} {\bibfield  {journal}
  {\bibinfo  {journal} {Phys. Rev. B}\ }\textbf {\bibinfo {volume} {92}},\
  \bibinfo {pages} {140412} (\bibinfo {year} {2015})}\BibitemShut {NoStop}%
\bibitem [{\citenamefont {Newman}\ and\ \citenamefont
  {Barkema}(1999)}]{Newman1999}%
  \BibitemOpen
  \bibfield  {author} {\bibinfo {author} {\bibfnamefont {M.~E.}\ \bibnamefont
  {Newman}}\ and\ \bibinfo {author} {\bibfnamefont {G.~T.}\ \bibnamefont
  {Barkema}},\ }\href@noop {} {\emph {\bibinfo {title} {Monte Carlo methods in
  statistical physics}}}\ (\bibinfo  {publisher} {Clarendon Press, Oxford,
  UK},\ \bibinfo {year} {1999})\BibitemShut {NoStop}%
\bibitem [{\citenamefont {White}(1992)}]{White1992}%
  \BibitemOpen
  \bibfield  {author} {\bibinfo {author} {\bibfnamefont {S.~R.}\ \bibnamefont
  {White}},\ }\bibfield  {title} {\bibinfo {title} {Density matrix formulation
  for quantum renormalization groups},\ }\href
  {https://doi.org/10.1103/PhysRevLett.69.2863} {\bibfield  {journal} {\bibinfo
   {journal} {Phys. Rev. Lett.}\ }\textbf {\bibinfo {volume} {69}},\ \bibinfo
  {pages} {2863} (\bibinfo {year} {1992})}\BibitemShut {NoStop}%
\bibitem [{\citenamefont {Gogolin}\ \emph {et~al.}(2004)\citenamefont
  {Gogolin}, \citenamefont {Nersesyan},\ and\ \citenamefont
  {Tsvelik}}]{Gogolin2004}%
  \BibitemOpen
  \bibfield  {author} {\bibinfo {author} {\bibfnamefont {A.~O.}\ \bibnamefont
  {Gogolin}}, \bibinfo {author} {\bibfnamefont {A.~A.}\ \bibnamefont
  {Nersesyan}},\ and\ \bibinfo {author} {\bibfnamefont {A.~M.}\ \bibnamefont
  {Tsvelik}},\ }\href@noop {} {\emph {\bibinfo {title} {Bosonization and
  strongly correlated systems}}}\ (\bibinfo  {publisher} {Cambridge University
  Press, U.K.},\ \bibinfo {year} {2004})\BibitemShut {NoStop}%
\bibitem [{\citenamefont {Chen}\ and\ \citenamefont
  {Nussinov}(2008)}]{Chen2008}%
  \BibitemOpen
  \bibfield  {author} {\bibinfo {author} {\bibfnamefont {H.-D.}\ \bibnamefont
  {Chen}}\ and\ \bibinfo {author} {\bibfnamefont {Z.}~\bibnamefont
  {Nussinov}},\ }\bibfield  {title} {\bibinfo {title} {Exact results of the
  {Kitaev} model on a hexagonal lattice: spin states, string and brane
  correlators, and anyonic excitations},\ }\href
  {https://doi.org/10.1088/1751-8113/41/7/075001} {\bibfield  {journal}
  {\bibinfo  {journal} {J. Phys. A: Math. Theor.}\ }\textbf {\bibinfo {volume}
  {41}},\ \bibinfo {pages} {075001} (\bibinfo {year} {2008})}\BibitemShut
  {NoStop}%
\bibitem [{\citenamefont {Yao}\ and\ \citenamefont {Oshikawa}(2020)}]{Yao2020}%
  \BibitemOpen
  \bibfield  {author} {\bibinfo {author} {\bibfnamefont {Y.}~\bibnamefont
  {Yao}}\ and\ \bibinfo {author} {\bibfnamefont {M.}~\bibnamefont {Oshikawa}},\
  }\bibfield  {title} {\bibinfo {title} {Generalized boundary condition applied
  to {Lieb-Schultz-Mattis-Type} ingappabilities and many-body chern numbers},\
  }\href {https://doi.org/10.1103/PhysRevX.10.031008} {\bibfield  {journal}
  {\bibinfo  {journal} {Phys. Rev. X}\ }\textbf {\bibinfo {volume} {10}},\
  \bibinfo {pages} {031008} (\bibinfo {year} {2020})}\BibitemShut {NoStop}%
\bibitem [{\citenamefont {Yamada}\ and\ \citenamefont
  {Fujimoto}(2022)}]{Yamada2022}%
  \BibitemOpen
  \bibfield  {author} {\bibinfo {author} {\bibfnamefont {M.~G.}\ \bibnamefont
  {Yamada}}\ and\ \bibinfo {author} {\bibfnamefont {S.}~\bibnamefont
  {Fujimoto}},\ }\bibfield  {title} {\bibinfo {title} {Thermodynamic signature
  of {SU(4)} spin-orbital liquid and symmetry fractionalization from
  lieb-schultz-mattis theorem},\ }\href
  {https://doi.org/10.1103/PhysRevB.105.L201115} {\bibfield  {journal}
  {\bibinfo  {journal} {Phys. Rev. B}\ }\textbf {\bibinfo {volume} {105}},\
  \bibinfo {pages} {L201115} (\bibinfo {year} {2022})}\BibitemShut {NoStop}%
\bibitem [{\citenamefont {LeBlanc}\ \emph {et~al.}(2015)\citenamefont
  {LeBlanc}, \citenamefont {Antipov}, \citenamefont {Becca}, \citenamefont
  {Bulik}, \citenamefont {Chan}, \citenamefont {Chung}, \citenamefont {Deng},
  \citenamefont {Ferrero}, \citenamefont {Henderson}, \citenamefont
  {Jim\'enez-Hoyos}, \citenamefont {Kozik}, \citenamefont {Liu}, \citenamefont
  {Millis}, \citenamefont {Prokof'ev}, \citenamefont {Qin}, \citenamefont
  {Scuseria}, \citenamefont {Shi}, \citenamefont {Svistunov}, \citenamefont
  {Tocchio}, \citenamefont {Tupitsyn}, \citenamefont {White}, \citenamefont
  {Zhang}, \citenamefont {Zheng}, \citenamefont {Zhu},\ and\ \citenamefont
  {Gull}}]{LeBlanc2015}%
  \BibitemOpen
  \bibfield  {author} {\bibinfo {author} {\bibfnamefont {J.~P.~F.}\
  \bibnamefont {LeBlanc}}, \bibinfo {author} {\bibfnamefont {A.~E.}\
  \bibnamefont {Antipov}}, \bibinfo {author} {\bibfnamefont {F.}~\bibnamefont
  {Becca}}, \bibinfo {author} {\bibfnamefont {I.~W.}\ \bibnamefont {Bulik}},
  \bibinfo {author} {\bibfnamefont {G.~K.-L.}\ \bibnamefont {Chan}}, \bibinfo
  {author} {\bibfnamefont {C.-M.}\ \bibnamefont {Chung}}, \bibinfo {author}
  {\bibfnamefont {Y.}~\bibnamefont {Deng}}, \bibinfo {author} {\bibfnamefont
  {M.}~\bibnamefont {Ferrero}}, \bibinfo {author} {\bibfnamefont {T.~M.}\
  \bibnamefont {Henderson}}, \bibinfo {author} {\bibfnamefont {C.~A.}\
  \bibnamefont {Jim\'enez-Hoyos}}, \bibinfo {author} {\bibfnamefont
  {E.}~\bibnamefont {Kozik}}, \bibinfo {author} {\bibfnamefont {X.-W.}\
  \bibnamefont {Liu}}, \bibinfo {author} {\bibfnamefont {A.~J.}\ \bibnamefont
  {Millis}}, \bibinfo {author} {\bibfnamefont {N.~V.}\ \bibnamefont
  {Prokof'ev}}, \bibinfo {author} {\bibfnamefont {M.}~\bibnamefont {Qin}},
  \bibinfo {author} {\bibfnamefont {G.~E.}\ \bibnamefont {Scuseria}}, \bibinfo
  {author} {\bibfnamefont {H.}~\bibnamefont {Shi}}, \bibinfo {author}
  {\bibfnamefont {B.~V.}\ \bibnamefont {Svistunov}}, \bibinfo {author}
  {\bibfnamefont {L.~F.}\ \bibnamefont {Tocchio}}, \bibinfo {author}
  {\bibfnamefont {I.~S.}\ \bibnamefont {Tupitsyn}}, \bibinfo {author}
  {\bibfnamefont {S.~R.}\ \bibnamefont {White}}, \bibinfo {author}
  {\bibfnamefont {S.}~\bibnamefont {Zhang}}, \bibinfo {author} {\bibfnamefont
  {B.-X.}\ \bibnamefont {Zheng}}, \bibinfo {author} {\bibfnamefont
  {Z.}~\bibnamefont {Zhu}},\ and\ \bibinfo {author} {\bibfnamefont
  {E.}~\bibnamefont {Gull}} (\bibinfo {collaboration} {Simons Collaboration on
  the Many-Electron Problem}),\ }\bibfield  {title} {\bibinfo {title}
  {{Solutions of the Two-Dimensional Hubbard Model: Benchmarks and Results from
  a Wide Range of Numerical Algorithms}},\ }\href
  {https://doi.org/10.1103/PhysRevX.5.041041} {\bibfield  {journal} {\bibinfo
  {journal} {Phys. Rev. X}\ }\textbf {\bibinfo {volume} {5}},\ \bibinfo {pages}
  {041041} (\bibinfo {year} {2015})}\BibitemShut {NoStop}%
\bibitem [{\citenamefont {Sandvik}\ and\ \citenamefont
  {Evertz}(2010)}]{Sandvik2010}%
  \BibitemOpen
  \bibfield  {author} {\bibinfo {author} {\bibfnamefont {A.~W.}\ \bibnamefont
  {Sandvik}}\ and\ \bibinfo {author} {\bibfnamefont {H.~G.}\ \bibnamefont
  {Evertz}},\ }\bibfield  {title} {\bibinfo {title} {Loop updates for
  variational and projector quantum monte carlo simulations in the valence-bond
  basis},\ }\href {https://doi.org/10.1103/PhysRevB.82.024407} {\bibfield
  {journal} {\bibinfo  {journal} {Phys. Rev. B}\ }\textbf {\bibinfo {volume}
  {82}},\ \bibinfo {pages} {024407} (\bibinfo {year} {2010})}\BibitemShut
  {NoStop}%
\bibitem [{\citenamefont {Nakamura}\ \emph {et~al.}(2021)\citenamefont
  {Nakamura}, \citenamefont {Masuda},\ and\ \citenamefont
  {Nishimoto}}]{Nakamura2021}%
  \BibitemOpen
  \bibfield  {author} {\bibinfo {author} {\bibfnamefont {M.}~\bibnamefont
  {Nakamura}}, \bibinfo {author} {\bibfnamefont {S.}~\bibnamefont {Masuda}},\
  and\ \bibinfo {author} {\bibfnamefont {S.}~\bibnamefont {Nishimoto}},\
  }\bibfield  {title} {\bibinfo {title} {Characterization of topological
  insulators based on the electronic polarization with spiral boundary
  conditions},\ }\href {https://doi.org/10.1103/PhysRevB.104.L121114}
  {\bibfield  {journal} {\bibinfo  {journal} {Phys. Rev. B}\ }\textbf {\bibinfo
  {volume} {104}},\ \bibinfo {pages} {L121114} (\bibinfo {year}
  {2021})}\BibitemShut {NoStop}%
\bibitem [{\citenamefont {Nishiyama}(2009)}]{Nishiyama2009}%
  \BibitemOpen
  \bibfield  {author} {\bibinfo {author} {\bibfnamefont {Y.}~\bibnamefont
  {Nishiyama}},\ }\bibfield  {title} {\bibinfo {title} {Deconfinement
  criticality in the spatially anisotropic triangular antiferromagnet with ring
  exchange},\ }\href {https://doi.org/10.1103/PhysRevB.79.054425} {\bibfield
  {journal} {\bibinfo  {journal} {Phys. Rev. B}\ }\textbf {\bibinfo {volume}
  {79}},\ \bibinfo {pages} {054425} (\bibinfo {year} {2009})}\BibitemShut
  {NoStop}%
\bibitem [{\citenamefont {Miyata}\ \emph {et~al.}(2021)\citenamefont {Miyata},
  \citenamefont {Hikihara}, \citenamefont {Furukawa}, \citenamefont {Kremer},
  \citenamefont {Zherlitsyn},\ and\ \citenamefont {Wosnitza}}]{Miyata2021}%
  \BibitemOpen
  \bibfield  {author} {\bibinfo {author} {\bibfnamefont {A.}~\bibnamefont
  {Miyata}}, \bibinfo {author} {\bibfnamefont {T.}~\bibnamefont {Hikihara}},
  \bibinfo {author} {\bibfnamefont {S.}~\bibnamefont {Furukawa}}, \bibinfo
  {author} {\bibfnamefont {R.~K.}\ \bibnamefont {Kremer}}, \bibinfo {author}
  {\bibfnamefont {S.}~\bibnamefont {Zherlitsyn}},\ and\ \bibinfo {author}
  {\bibfnamefont {J.}~\bibnamefont {Wosnitza}},\ }\bibfield  {title} {\bibinfo
  {title} {Magnetoelastic study on the frustrated quasi-one-dimensional
  spin-$\frac{1}{2}$ magnet {${\mathrm{LiCuVO}}_{4}$}},\ }\href
  {https://doi.org/10.1103/PhysRevB.103.014411} {\bibfield  {journal} {\bibinfo
   {journal} {Phys. Rev. B}\ }\textbf {\bibinfo {volume} {103}},\ \bibinfo
  {pages} {014411} (\bibinfo {year} {2021})}\BibitemShut {NoStop}%
\bibitem [{\citenamefont {White}\ and\ \citenamefont
  {Chernyshev}(2007)}]{White2007}%
  \BibitemOpen
  \bibfield  {author} {\bibinfo {author} {\bibfnamefont {S.~R.}\ \bibnamefont
  {White}}\ and\ \bibinfo {author} {\bibfnamefont {A.~L.}\ \bibnamefont
  {Chernyshev}},\ }\bibfield  {title} {\bibinfo {title} {{Ne\'el Order in
  Square and Triangular Lattice Heisenberg Models}},\ }\href
  {https://doi.org/10.1103/PhysRevLett.99.127004} {\bibfield  {journal}
  {\bibinfo  {journal} {Phys. Rev. Lett.}\ }\textbf {\bibinfo {volume} {99}},\
  \bibinfo {pages} {127004} (\bibinfo {year} {2007})}\BibitemShut {NoStop}%
\bibitem [{\citenamefont {Weihong}\ \emph
  {et~al.}(1991{\natexlab{a}})\citenamefont {Weihong}, \citenamefont {Oitmaa},\
  and\ \citenamefont {Hamer}}]{Weihong1991-1}%
  \BibitemOpen
  \bibfield  {author} {\bibinfo {author} {\bibfnamefont {Z.}~\bibnamefont
  {Weihong}}, \bibinfo {author} {\bibfnamefont {J.}~\bibnamefont {Oitmaa}},\
  and\ \bibinfo {author} {\bibfnamefont {C.~J.}\ \bibnamefont {Hamer}},\
  }\bibfield  {title} {\bibinfo {title} {Square-lattice {Heisenberg}
  antiferromagnet at {$T=0$}},\ }\href
  {https://doi.org/10.1103/PhysRevB.43.8321} {\bibfield  {journal} {\bibinfo
  {journal} {Phys. Rev. B}\ }\textbf {\bibinfo {volume} {43}},\ \bibinfo
  {pages} {8321} (\bibinfo {year} {1991}{\natexlab{a}})}\BibitemShut {NoStop}%
\bibitem [{\citenamefont {Hamer}\ \emph {et~al.}(1992)\citenamefont {Hamer},
  \citenamefont {Weihong},\ and\ \citenamefont {Arndt}}]{Hamer1992}%
  \BibitemOpen
  \bibfield  {author} {\bibinfo {author} {\bibfnamefont {C.~J.}\ \bibnamefont
  {Hamer}}, \bibinfo {author} {\bibfnamefont {Z.}~\bibnamefont {Weihong}},\
  and\ \bibinfo {author} {\bibfnamefont {P.}~\bibnamefont {Arndt}},\ }\bibfield
   {title} {\bibinfo {title} {Third-order spin-wave theory for the {Heisenberg}
  antiferromagnet},\ }\href {https://doi.org/10.1103/PhysRevB.46.6276}
  {\bibfield  {journal} {\bibinfo  {journal} {Phys. Rev. B}\ }\textbf {\bibinfo
  {volume} {46}},\ \bibinfo {pages} {6276} (\bibinfo {year}
  {1992})}\BibitemShut {NoStop}%
\bibitem [{\citenamefont {Lin}\ \emph {et~al.}(2001)\citenamefont {Lin},
  \citenamefont {Flynn},\ and\ \citenamefont {Betts}}]{Lin2001}%
  \BibitemOpen
  \bibfield  {author} {\bibinfo {author} {\bibfnamefont {H.-Q.}\ \bibnamefont
  {Lin}}, \bibinfo {author} {\bibfnamefont {J.~S.}\ \bibnamefont {Flynn}},\
  and\ \bibinfo {author} {\bibfnamefont {D.~D.}\ \bibnamefont {Betts}},\
  }\bibfield  {title} {\bibinfo {title} {Exact diagonalization and quantum
  {Monte Carlo} study of the spin-$\frac{1}{2}$ $\mathrm{XXZ}$ model on the
  square lattice},\ }\href {https://doi.org/10.1103/PhysRevB.64.214411}
  {\bibfield  {journal} {\bibinfo  {journal} {Phys. Rev. B}\ }\textbf {\bibinfo
  {volume} {64}},\ \bibinfo {pages} {214411} (\bibinfo {year}
  {2001})}\BibitemShut {NoStop}%
\bibitem [{\citenamefont {Bishop}\ \emph {et~al.}(2017)\citenamefont {Bishop},
  \citenamefont {Li}, \citenamefont {Zinke}, \citenamefont {Darradi},
  \citenamefont {Richter}, \citenamefont {Farnell},\ and\ \citenamefont
  {Schulenburg}}]{Bishop2017}%
  \BibitemOpen
  \bibfield  {author} {\bibinfo {author} {\bibfnamefont {R.}~\bibnamefont
  {Bishop}}, \bibinfo {author} {\bibfnamefont {P.}~\bibnamefont {Li}}, \bibinfo
  {author} {\bibfnamefont {R.}~\bibnamefont {Zinke}}, \bibinfo {author}
  {\bibfnamefont {R.}~\bibnamefont {Darradi}}, \bibinfo {author} {\bibfnamefont
  {J.}~\bibnamefont {Richter}}, \bibinfo {author} {\bibfnamefont
  {D.}~\bibnamefont {Farnell}},\ and\ \bibinfo {author} {\bibfnamefont
  {J.}~\bibnamefont {Schulenburg}},\ }\bibfield  {title} {\bibinfo {title} {The
  spin-half {XXZ} antiferromagnet on the square lattice revisited: A high-order
  coupled cluster treatment},\ }\href
  {https://doi.org/https://doi.org/10.1016/j.jmmm.2016.11.043} {\bibfield
  {journal} {\bibinfo  {journal} {J. Magn. Magn. Mater.}\ }\textbf {\bibinfo
  {volume} {428}},\ \bibinfo {pages} {178} (\bibinfo {year}
  {2017})}\BibitemShut {NoStop}%
\bibitem [{\citenamefont {Weihong}\ \emph
  {et~al.}(1991{\natexlab{b}})\citenamefont {Weihong}, \citenamefont {Oitmaa},\
  and\ \citenamefont {Hamer}}]{Weihong1991-2}%
  \BibitemOpen
  \bibfield  {author} {\bibinfo {author} {\bibfnamefont {Z.}~\bibnamefont
  {Weihong}}, \bibinfo {author} {\bibfnamefont {J.}~\bibnamefont {Oitmaa}},\
  and\ \bibinfo {author} {\bibfnamefont {C.~J.}\ \bibnamefont {Hamer}},\
  }\bibfield  {title} {\bibinfo {title} {Second-order spin-wave results for the
  quantum {XXZ} and {XY} models with anisotropy},\ }\href
  {https://doi.org/10.1103/PhysRevB.44.11869} {\bibfield  {journal} {\bibinfo
  {journal} {Phys. Rev. B}\ }\textbf {\bibinfo {volume} {44}},\ \bibinfo
  {pages} {11869} (\bibinfo {year} {1991}{\natexlab{b}})}\BibitemShut {NoStop}%
\bibitem [{\citenamefont {Huse}(1988)}]{Huse1988}%
  \BibitemOpen
  \bibfield  {author} {\bibinfo {author} {\bibfnamefont {D.~A.}\ \bibnamefont
  {Huse}},\ }\bibfield  {title} {\bibinfo {title} {Ground-state staggered
  magnetization of two-dimensional quantum {Heisenberg} antiferromagnets},\
  }\href {https://doi.org/10.1103/PhysRevB.37.2380} {\bibfield  {journal}
  {\bibinfo  {journal} {Phys. Rev. B}\ }\textbf {\bibinfo {volume} {37}},\
  \bibinfo {pages} {2380} (\bibinfo {year} {1988})}\BibitemShut {NoStop}%
\bibitem [{\citenamefont {Kadosawa}\ \emph {et~al.}(2023)\citenamefont
  {Kadosawa}, \citenamefont {Nakamura}, \citenamefont {Ohta},\ and\
  \citenamefont {Nishimoto}}]{Kadosawa2023}%
  \BibitemOpen
  \bibfield  {author} {\bibinfo {author} {\bibfnamefont {M.}~\bibnamefont
  {Kadosawa}}, \bibinfo {author} {\bibfnamefont {M.}~\bibnamefont {Nakamura}},
  \bibinfo {author} {\bibfnamefont {Y.}~\bibnamefont {Ohta}},\ and\ \bibinfo
  {author} {\bibfnamefont {S.}~\bibnamefont {Nishimoto}},\ }\bibfield  {title}
  {\bibinfo {title} {Study of staggered magnetization in the spin-{$S$}
  square-lattice {Heisenberg} model using spiral boundary conditions},\ }\href
  {https://doi.org/10.7566/JPSJ.92.023701} {\bibfield  {journal} {\bibinfo
  {journal} {J. Phys. Soc. Jpn.}\ }\textbf {\bibinfo {volume} {92}},\ \bibinfo
  {pages} {023701} (\bibinfo {year} {2023})}\BibitemShut {NoStop}%
\end{thebibliography}%

\end{document}


\title{Supplementary Material for\\ ``One-dimensional projection of two-dimensional systems using spiral boundary conditions''}

\author{Masahiro Kadosawa$^1$, Masaaki Nakamura$^2$, Yukinori Ohta$^1$, and Satoshi Nishimoto$^{3,4}$}

\address{
$^1$ Department of Physics, Chiba University, Chiba 263-8522, Japan\\
$^2$ Department of Physics, Ehime University, Ehime 790-8577, Japan\\
$^3$ Department of Physics, Technical University Dresden, 01069 Dresden, Germany\\
$^4$ Institute for Theoretical Solid State Physics, IFW Dresden, 01069 Dresden, Germany
}

\date{\today}

\begin{abstract}
Here we present a detailed analysis of the ground-state energy of the
half-filled tight-binding model on a square lattice using spiral boundary
conditions. We also discuss the difference in the density of states between
using periodic and spiral boundary conditions. In addition, the detailed
data of our DMRG calculations with spiral boundary conditions are compared
to those with various boundary conditions. Finally, possible advantage of
spiral boundary conditions in DMRG simulation is discussed.
\end{abstract}

\maketitle

\section{Ground-state energy of the tight-binding model on a square lattice}

Using periodic boundary conditions (PBC), the Hamiltonian of the tight-binding
model on a square lattice is written as
\begin{equation}\label{ham2D}
{\cal H}_{0,\rm sq} =-t \sum_{i,\sigma}(c^\dagger_{(x,y),\sigma}c_{(x+1,y),\sigma}+c^\dagger_{(x,y),\sigma}c_{(x,y+1),\sigma}+{\rm H.c.})
\end{equation}
where $c_{(x,y),\sigma}$ is an annihilation operator of electron with spin
$\sigma$ at site $(x,y)$, $t$ is nearest-neighbor hopping integral, and
$c_{(x,y),\sigma}=c_{(x+n_xL_x,y+n_yL_y),\sigma}$, where $n_x$ and $n_y$ are
integers. The Fourier transform of \eqref{ham2D} reads
\begin{equation}\label{hamk2D}
{\cal H}_{{\rm sq},{\bf k}} =-2t \sum_{{\bf k},\sigma}(\cos k_x + \cos k_y) c^\dagger_{{\bf k},\sigma}c_{{\bf k},\sigma}
\end{equation}
where $c_{{\bf k},\sigma}=\frac{1}{\sqrt{L_xL_y}}\sum_{x,y} \exp(i{\bf k}\cdot {\bf r})c_{(x,y),\sigma}$ with ${\bf r}=(x,y)$ and
${\bf k}=(k_x,k_y)=\left(\frac{2\pi}{L_x}m_x,\frac{2\pi}{L_y}m_y\right)$ [$m_x=0,1,\dots,L_x-1; m_y=0,1,\dots,L_y-1$]. In the thermodynamic limit
the ground-state energy at half filling is given by
\begin{equation}\label{ene2D}
e_0=\frac{-2t}{(2\pi)^2}\int\int_{\varepsilon(K)<0}dk_xdk_y(\cos k_x + \cos k_y)=-\frac{16}{\pi^2}t.
\end{equation}

\begin{figure}[h]
	\includegraphics[width=0.5\linewidth]{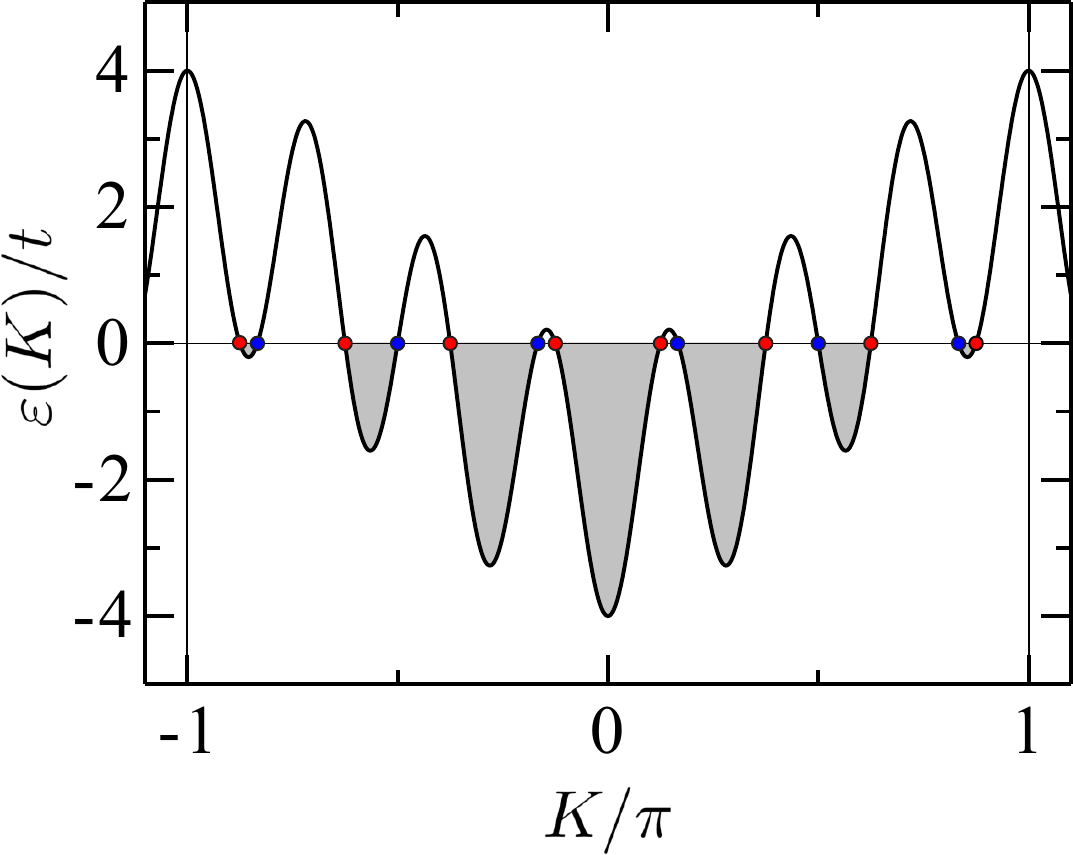}
	\caption{Energy dispersion of the tight-bonding model on a square lattice
		with spiral boundary conditions, where $L_x=8$ and $L_y\to\infty$ are
		taken. At half filling the shaded areas are occupied by electrons.
	}
	\label{fig:ek_SM}
\end{figure}

As explained in the main text, using spiral boundary conditions (SBC) the
original two-dimensional (2D) square-lattice cluster with $L_x \times L_y$ sites
can be exactly reproduced as a one-dimensional (1D) chain with nearest- and $(L_x-1)$-th neighbor hopping integrals, where the translational symmetry is preserved. The Hamiltonian is written as
\begin{equation}\label{ham1D}
{\cal H}_{0,\rm sq} =-t \sum_\sigma\sum_{i=1}^{L_xL_y}(c^\dagger_{i,\sigma}c_{i+1,\sigma}+c^\dagger_{i,\sigma}c_{i+(L_x-1),\sigma}+{\rm H.c.}),
\end{equation}
where $c_{i,\sigma}$ is an annihilation operator of electron with spin $\sigma$
at site $i$. Its Fourier transform leads to
\begin{equation}\label{hamk1D}
{\cal H}_{{\rm sq},K} =-2t \sum_{K,\sigma}(\cos K + \cos (L_x-1)K) c^\dagger_{K,\sigma}c_{K,\sigma},
\end{equation}
where $c_{K,\sigma}=(1/\sqrt{L_xL_y})\sum_i \exp(iKr_i)c_{i,\sigma}$,
and the momentum $K$ is defined along the projected 1D periodic chain as
$K=\frac{2\pi}{L_xL_y}n$ with $n=n_x+L_xn_y=0,1,\dots,L_xL_y-1)$.

The energy dispersion \eqref{hamk1D} for $L_x=8$ and $L_y\to\infty$ is
shown in Fig.~\ref{fig:ek_SM}. Let us then calculate the ground-state
energy at half filling, where the shaded areas are occupied by electrons.
Since $\cos K + \cos (L_x-1)K=2\cos\frac{L_xK}{2}\cos\frac{(L_x-2)K}{2}$,
the Fermi points denoted by red and blue circles are given as
\begin{align}
\frac{L_xK_r}{2}=\frac{\pi}{2}+n_r\pi &\longrightarrow K_r=\frac{\pi}{L_x}(2n_r-1)\\
\frac{(L_x-2)K_b}{2}=\frac{\pi}{2}+n_b\pi &\longrightarrow K_b=\frac{\pi}{L_x-2}(2n_b-1),
\end{align}
respectively, where $n_r,n_b=1,2,\cdots,$
Thus,
\begin{align}\label{ene1D}
\nonumber
e_0=\frac{-2t}{\pi}&\int_{\varepsilon(K)<0}dK(\cos K + \cos (L_x-1)K)\\
\nonumber
=\frac{-4t}{\pi}&\int_0^\pi dK(\cos K + \cos (L_x-1)K)\theta(-\varepsilon(K))\\
\nonumber
=\frac{-4t}{\pi}&\left(\left[\sin K+\frac{\sin[(L_x-1)K]}{L_x-1}\right]_0^\frac{\pi}{L_x}
+\left[\sin K+\frac{\sin[(L_x-1)K]}{L_x-1}\right]_\frac{\pi}{L_x-2}^\frac{3\pi}{L_x}
+\cdots+\left[\sin K+\frac{\sin[(L_x-1)K]}{L_x-1}\right]_\frac{(L_x-3)\pi}{L_x-2}^\frac{(L_x-1)\pi}{L_x}\right)\\
\nonumber
=\frac{-4t}{\pi}&\left[
\left(\sin\frac{\pi}{L_x}+\sin\frac{3\pi}{L_x}+\cdots+\sin\frac{(L_x-1)\pi}{L_x}\right)
+\frac{1}{L_x-1}\left(\sin\frac{(L_x-1)\pi}{L_x}+\sin\frac{3(L_x-1)\pi}{L_x}+\cdots+\sin\frac{(L_x-1)^2\pi}{L_x}\right) \right. \\
\nonumber
&\left.-\left(\sin\frac{\pi}{L_x-2}+\sin\frac{3\pi}{L_x-2}+\cdots+\sin\frac{(L_x-3)\pi}{L_x-2}\right) \right. \\
&\left.-\frac{1}{L_x-1}\left(\sin\frac{(L_x-1)\pi}{L_x-2}+\sin\frac{3(L_x-1)\pi}{L_x-2}+\cdots+\sin\frac{(L_x-3)(L_x-1)\pi}{L_x-2}\right)\right].
\end{align}
Applying the formula
$\sin x+\sin (3x)+\cdots+\sin[(2n-1)x]=\frac{1-\cos (2nx)}{2\sin x}$
to Eq.~\eqref{ene1D}, we obtain
\begin{align}
e_0&=\frac{-4t}{\pi}\left(\frac{1}{\sin\frac{\pi}{L_x}}+\frac{1}{(L_x-1)\sin\left(\frac{L_x-1}{L_x}\pi\right)}-\frac{1}{\sin\frac{\pi}{L_x-2}}-\frac{1}{(L_x-1)\sin\left(\frac{L_x-1}{L_x-2}\pi\right)}\right)\\
&\underset{L_x\gg1}{\longrightarrow}\frac{-4t}{\pi}\left(\frac{L_x}{\pi}+\frac{L_x}{\pi(L_x-1)}-\frac{L_x-2}{\pi}+\frac{L_x-2}{\pi(L_x-1)}\right)
\underset{L_x\to\infty}{\longrightarrow}-\frac{16}{\pi^2}.
\end{align}
This coincides with Eq.~\eqref{ene2D}. Thus, we can confirm that the
finite-size systems under SBC are adiabatically connected to the thermodynamic
limit.

\newpage

\section{Splitting of the degenerate energy levels by spiral boundary conditions}

\begin{figure}[h]
	\includegraphics[width=0.6\linewidth]{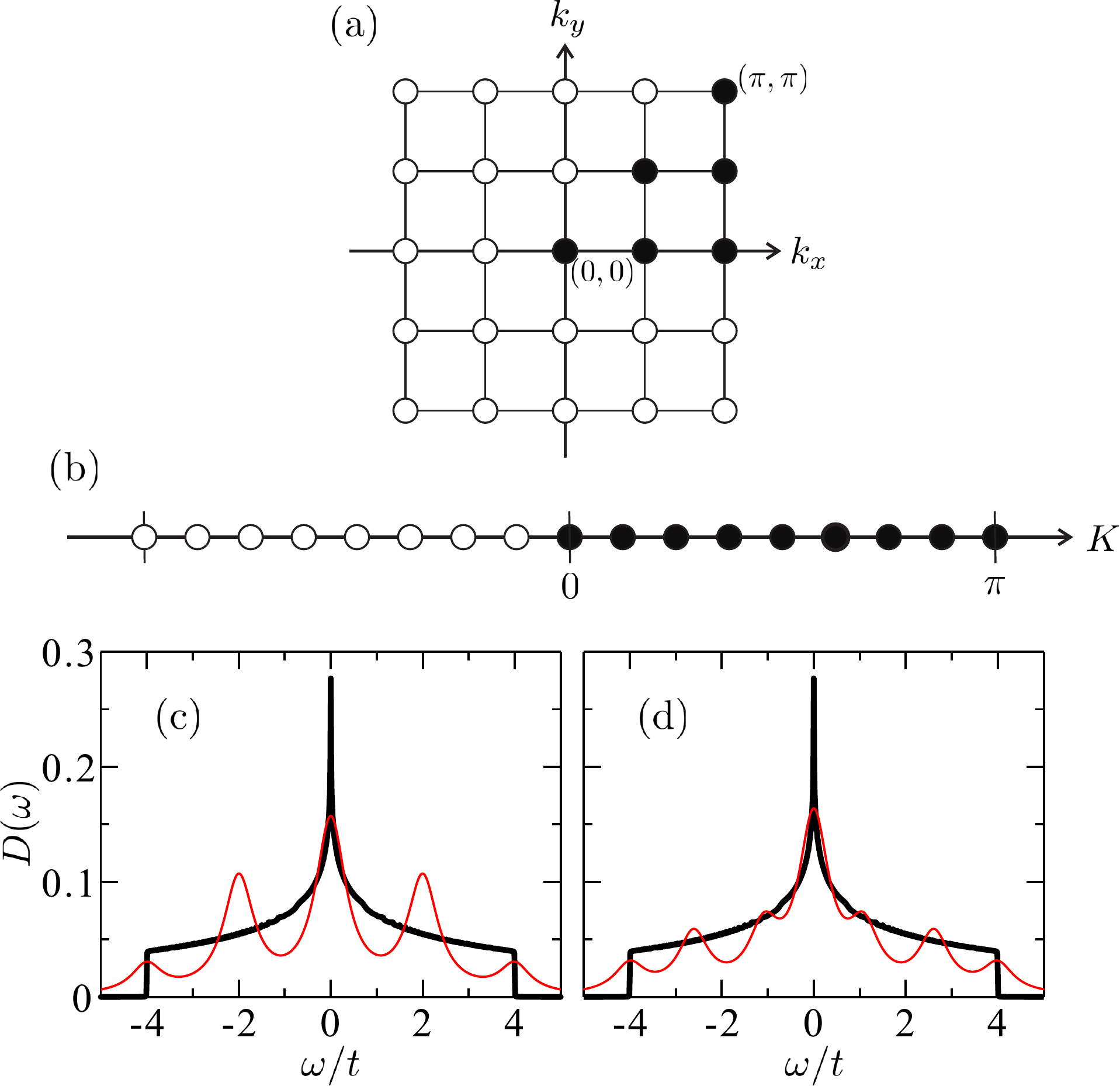}
	\caption{(a,b) Possible momenta for a $4 \times 4$ square-lattice cluster
		under periodic and spiral boundary conditions, respectively.
		Example sets of independent momenta are denoted by filled circles.
		(c,d) Density of states for the square-lattice tight-binding model
		with a $4 \times 4$ sites with periodic and spiral boundary conditions,
		respectively.
	}
	\label{fig:DOS_SM}
\end{figure}

The symmetries of cluster are different between using SBC and periodic 
boundary conditions (PBC). Let us here explain it by taking a square-lattice
cluster as an example. A cluster with $L \times L$ sites has a $C_4$ rotational
symmetry under PBC. Accordingly, the number of independent momenta is
$\frac{L^2}{8}+\frac{3L}{4}+1$. On the other hand, when SBC are applied,
the rotational symmetry is reduced from $C_4$ to $C_2$ due to the splitting
of the degenerate energy levels. Then, the number of independent momenta
is increased to $\frac{L^2}{2}+1$. In Fig.~\ref{fig:DOS_SM}(a,b), possible
momenta for a $4 \times 4$ cluster under PBC and SBC are shown, and
the independent ones denoted by filled circles are $6$ and $9$, respectively.
As a result, the density of states with SBC is closer to the thermodynamic
limit one, see Fig.~\ref{fig:DOS_SM}(c,d).

\newpage

\section{Flexibility of spiral boundary conditions}

\begin{figure}[h]
	\includegraphics[width=1.0\linewidth]{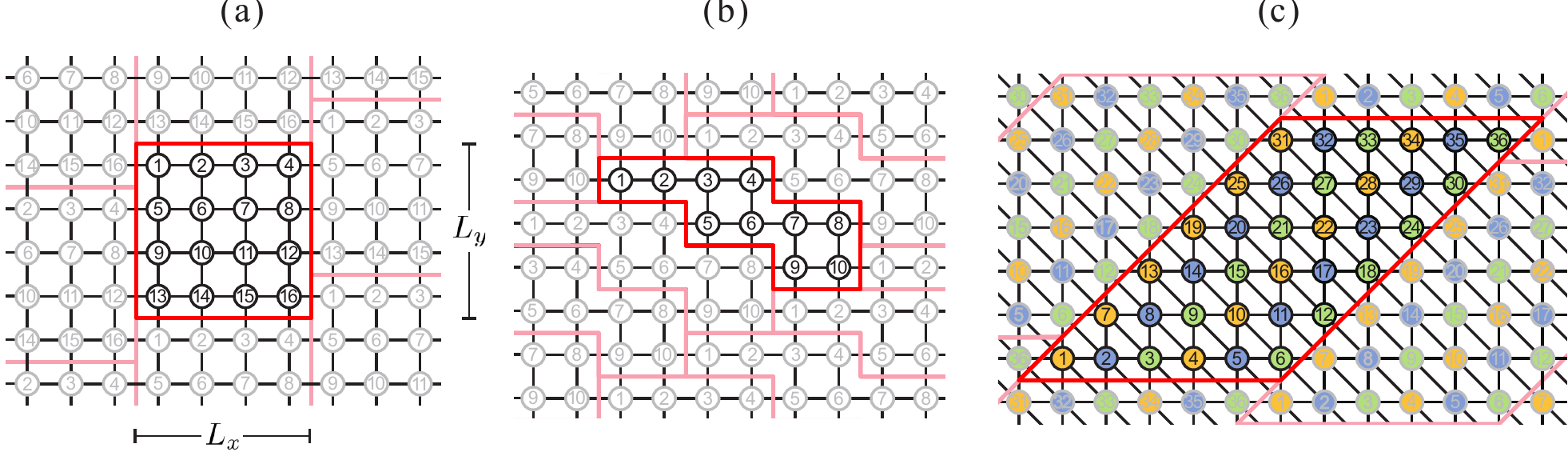}
	\caption{(a) 2D square-lattice cluster where a region framed by red line is the original cluster. This application of SBC is consistent with an ordered state with modulation of $\bm{k}=(\pi,0)$. (b) A possible application of
    SBC for 10-site non-rectangular cluster. (c) 2D triangular-lattice cluster
    where a region framed by red line is the original cluster. This application
    of SBC is consistent with an ordered state with modulation of
    $\bm{k}=(2\pi/3,2\pi/3)$. The sites belonging to different sublattices are
    color-coded.
	}
	\label{fig:anothersbc_SM}
\end{figure}

As noted in the main text, the way of 1D projection using SBC is not unique~\cite{Nakamura}.
This means that the modulation of wave function can be controlled more flexibly
than PBC. The way used for 2D square lattice in the main text is appropriate to
study a state with modulation of $\bm{k}=(\pi,\pi)$, like a N\'eel state.
However, if we study a state with  modulation of $\bm{k}=(\pi,0)$, it is more
convenient to use another application way of SBC. It is shown in
Fig.~\ref{fig:anothersbc_SM}(a). In this way, the original 2D square-lattice
cluster with $L_x \times L_y$ sites can be exactly mapped onto a 1D chain with
nearest- and $L_x$-th neighbor hopping integrals, where the translation symmetry is also preserved. The Hamiltonian is written as
\begin{equation}
	{\cal H}_{0,\rm sq} =-t \sum_\sigma\sum_{i=1}^{L_xL_y}(c^\dagger_{i,\sigma}c_{i+1,\sigma}+c^\dagger_{i,\sigma}c_{i+L_x,\sigma}+{\rm H.c.}),
\end{equation}
and its Fourier transform leads to
\begin{equation}
	{\cal H}_{{\rm sq},K} =-2t \sum_{K,\sigma}(\cos K + \cos L_xK)c^\dagger_{K,\sigma}c_{K,\sigma},
\end{equation}
where $n=n_x+L_xn_y=0,1,\dots,L_xL_y-1)$. It is interesting that both $\bm{k}=(\pi,\pi)$ in Fig.~1(a) of the main text and 
$\bm{k}=(\pi,0)$ in Fig.~\ref{fig:anothersbc_SM}(a) are 
represented by a $k=\pi$ modulation in the projected 1D chain.

In fact, when we apply SBC, there is in general no reason to stick to 
lattices which are square, rectangule, or parallelogram. A cluster shape
and number of sites can be relatively-flexibly chosen under SBC.
For instance, we can construct a lattice with 10 sites as shown in
Fig.~\ref{fig:anothersbc_SM}(b). We can easily confirm that the mapped
1D chain has nearest- and 2nd neighbor bonds and the translational
symmetry is preserved. Thus, we are allowed to control the application way
of SBC depending on considered physics.

As another example, let us consider a state with modulation of
$\bm{k}=(2\pi/3,2\pi/3)$. This corresponds to a 120$^\circ$
structure in the triangular-lattice Heisenberg model. Similarly,
this state can be represented by 1D chain with translation symmetry,
as shown in Fig.~\ref{fig:anothersbc_SM}(c). For a cluster with
$L_x \times L_y$ sites, the 2D system can be exactly mapped onto
a 1D chain with nearest-, $(L_x-2)$-th, and $(L_x-1)$-th neighbor bonds.

As shown above, the modulation of cluster can be flexibly controlled.
In the case of incommensurate order, it would be able to make
a convergence of physical quantity to the thermodynamic limit faster
by realizing a possibly closest modulation to the incommensurate vector.

\newpage

\section{Correlation functions for various boundary conditions}

\begin{figure}[h]
	\includegraphics[width=0.8\linewidth]{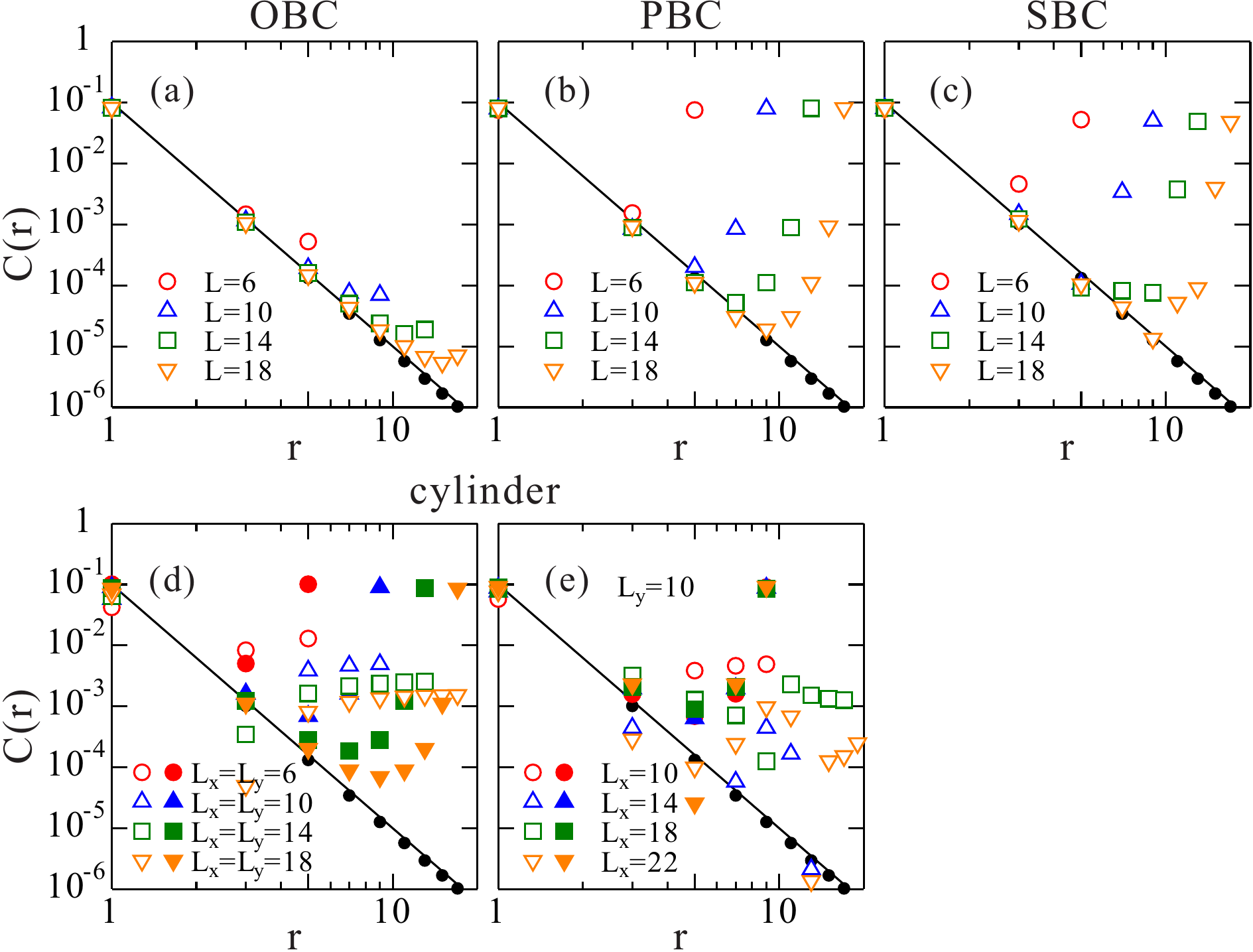}
	\caption{Density-density correlation functions $C(r)$ for non-interacting
		fermions on a 2D square lattice at half filling as a function of
		distance $r$ along bond direction, namely, $(1,0)$ or $(0,1)$, with
		(a) OBC, (b) PBC, (c) SBC, and (d,e) cylinder. For cylinder,
		the values for both $(1,0)$ and $(0,1)$ directions are shown.
		As a reference data, the result for very large system with $100\times100$ OBC cluster is plotted by black dots. As expected,
		it decays like $C(r) \propto 1/r^4$ (black line).
	}
	\label{fig:U0corr_SM}
\end{figure}

It is interesting to see how correlation functions are affected by boundary
conditions. To investigate their long-range behavior, we consider the
density-density correlation functions $C(r)=1-\langle n_i n_{i+r} \rangle$
of non-interacting fermions on a 2D square lattice because they can be
exactly obtained. In Fig.~\ref{fig:U0corr_SM} we plot
$C(r)=1-\langle n_i n_{i+r} \rangle$ as a function of distance $r$
along bond direction, namely, $(1,0)$ or $(0,1)$, at half filling for
various boundary conditions. We find that OBC, PBC, and SBC can provide
equally good results: For example, the data points up to $r\sim9$ are
almost on the analytical line $C(r) \propto 1/r^4$ for the systems with
$L=18$. On the other hand, the result with cylindrical cluster is obviously
worst, namely, the deviation from $C(r) \propto 1/r^4$ is largest.
As is well known~\cite{Sandvik}, the result seems to be improved
by increasing the system size with keeping the ratio between system length
and circumference. Nevertheless, the behavior $C(r) \propto 1/r^4$
can be confirmed only up to $r\sim5$ in the case of $L_x=L_y=18$
[see Fig.~\ref{fig:U0corr_SM}(d)]. It is because that cylindrical boundary
conditions impose a kind of spatial anisotropy on the original isotropic
2D square-lattice cluster. Furthermore, surprisingly, the result cannot
be improved by increasing the system length if the circumference is kept
[Fig.~\ref{fig:U0corr_SM}(e)].

As discussed above, we may choose either OBC or SBC when correlation functions
for 2D system are studied. However, a 2D system under OBC can be easily
disturbed by Friedel oscillations once it is doped or any interaction is
switched on. Therefore, SBC could be a first choice for DMRG study of correlation functions for 2D systems.

\newpage

\section{Size scaling of ground-state energy for various boundary conditions}

\begin{figure}[h]
	\includegraphics[width=0.8\linewidth]{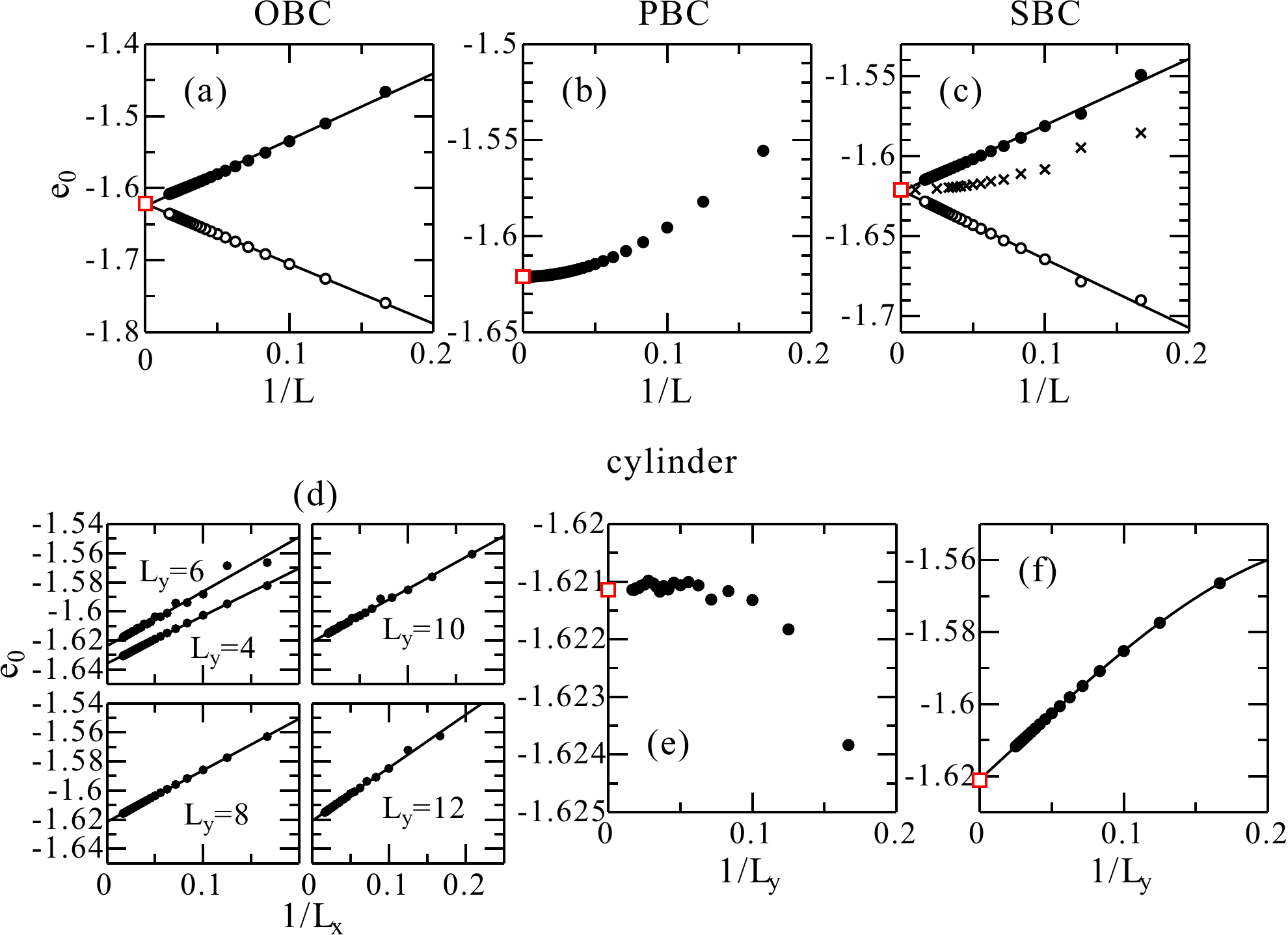}
	\caption{Finite-size-scaling analysis of the ground-state energy for
		non-interacting fermions on a 2D square lattice at half filling
		with (a) OBC, (b) PBC, (c) SBC, and (d-f) cylinder. The exact
		ground-state energy in the thermodynamic limit, $e_0=-16/\pi^2$,
		is denoted by red square. The solid and open circles denote the
		ground state energy per site and per two bonds, respectively.
		For cylinder, scaling analyses along system length $L_x$ for given
		circumferences $L_y$ are performed in (d), and then scaling analysis
		along circumference direction is performed in (e), where the
		extrapolated values obtained in (d) are used. Also, scaling analysis
		with keeping $L_x=L_y$ is performed in (f).
	}
	\label{fig:U0energy_SM}
\end{figure}

We here look into finite-size-scaling behavior of the ground-state energy
for each of boundary conditions. In Fig.~\ref{fig:U0energy_SM} we perform
finite-size-scaling analyses of the ground-state energy for non-interacting
fermions on a 2D square lattice at half filling with various boundary
conditions. Note that the result for SBC [Fig.~\ref{fig:U0energy_SM}(c)]
is equivalent to Fig.~3(c) in the main text. The scaling behavior for OBC
is straightforward [Fig.~\ref{fig:U0energy_SM}(a)], which is similar to
that for the case of 1D projected open chain under SBC
[Fig.~\ref{fig:U0energy_SM}(c)]. On the other hand, the ground-state energy
approaches to its thermodynamic-limit value as $1/L^2$ for PBC. If we use
DMRG, the ground-state energies can be calculated only for $L\lesssim10$
with 2D periodic clusters. Thus, the finite-size-scaling analysis of the
ground-state energy would be difficult for PBC.

When we use cylinder, finite-size-scaling analysis must be performed along
two orientations, e.g., $x$- and $y$-directions. First, the scaling analyses
along system length $L_x$ for given circumferences $L_y$ are performed to
obtain the $L_x\to\infty$ values of the ground-state energy for each $L_y$
[Fig.~\ref{fig:U0energy_SM}(d)]. Next, with using the obtained $L_x\to\infty$
values further scaling analysis along circumference direction $L_y$ is
performed [Fig.~\ref{fig:U0energy_SM}(e)]. Similarly to the case of PBC,
the scaling behavior looks like $e_0=-16/\pi^2+\alpha/L_y^\beta$
($\alpha<0$, $\beta>2$). Therefore, accurate estimation of the ground-state
energy for 2D itinerant systems using DMRG with cylindrical clusters may
be not very practical, considering that there would be some uncertainty
in the extrapolation to $L_x\to\infty$ and the extrapolated values can be
obtained only for $L_y\lesssim10$. We also perform a scaling analysis with
keeping $L_x=L_y$ in Fig.~\ref{fig:U0energy_SM}(f). The scaling behavior
looks simple but deviates from a linear function like for OBC and open chain
under SBC.

For the above reason, OBC or SBC would be a best choice for estimating the
ground-state energy of 2D systems in the thermodynamic limit.

\newpage

\section{Convergence check of DMRG calculations}

\begin{table}[tbh]
	\caption{Discarded weight $w_d$ and the ground-state energy $E_0(m)$
		 of the half-filled square-lattice Hubbard model at $U/t=8$ as
		 functions of system size $L \times L$ and the number of
		 density-matrix eigenstates kept in the renormalization $m$,
		 where SBC are applied. The ground-state energy at $m\to\infty$
		 was obtained by a linear fit of $E_0(m)$ vs. $w_d$.
	 }
    \begin{minipage}[t]{.48\textwidth}
	\begin{center}
	\begin{tabular}{wc{1.5cm}wc{1.8cm}wc{2cm}wc{3cm}}
		\multicolumn{4}{c}{$L=6$} \\
		\hline
		\hline
		$m$ &  $w_d$  & $E_0(m)$ &  $E_0(m)-E_0(\infty)$ \\
		\hline
		2000  &   3.432e-06  &  -17.79364790  &  0.01227902 \\
		4000  &   7.856e-07  &  -17.80326286  &  0.00266406 \\
		6000  &   3.052e-07  &  -17.80489349  &  0.00103344 \\
		8000  &   1.516e-07  &  -17.80541043  &  0.00051649 \\
		10000  &   8.614e-08  &  -17.80563244  &  0.00029448 \\
		12000  &   5.332e-08  &  -17.80574150  &  0.00018542 \\
		\hline
		$\infty$ &             &   -17.80592692  &            \\
		\hline
		\hline
	\end{tabular}
	\end{center}
    \end{minipage}
\hfill
    \begin{minipage}[t]{.48\textwidth}
	\begin{center}
	\begin{tabular}{wc{1.5cm}wc{1.8cm}wc{2cm}wc{3cm}}
	\multicolumn{4}{c}{$L=8$} \\
	\hline
	\hline
	$m$ &  $w_d$  & $E_0(m)$ &  $E_0(m)-E_0(\infty)$ \\
	\hline
2000   &  1.286e-05  &  -31.76311291   &   0.26322934 \\
4000   &   8.420e-06  &  -31.87302879   &   0.15331346 \\	
6000   &  5.681e-06  &  -31.91448613   &   0.11185612 \\
8000   &   4.550e-06  &  -31.93686464   &   0.08947761 \\
10000   &  4.165e-06  &  -31.95093410   &   0.07540815 \\
12000   &  3.815e-06  &  -31.96253340   &   0.06380885 \\
	\hline
	$\infty$ &             &   -32.02634225  &            \\
	\hline
	\hline
	\end{tabular}
	\end{center}
	\end{minipage}
\hfill
    \begin{minipage}[t]{.48\textwidth}
	\begin{center}
		\begin{tabular}{wc{1.5cm}wc{1.8cm}wc{2cm}wc{3cm}}
			\multicolumn{4}{c}{} \\
			\multicolumn{4}{c}{$L=10$} \\
			\hline
			\hline
			$m$ &  $w_d$  & $E_0(m)$ &  $E_0(m)-E_0(\infty)$ \\
			\hline
 2000   &  2.119e-05  &  -49.50077175  &  1.02132882 \\
4000   &  1.379e-05  &  -49.84727132  &  0.67482925 \\
6000   &  1.198e-05  &  -49.99628942  &  0.52581115 \\
8000   &  9.848e-06  &  -50.09485526  &  0.42724531 \\
10000   &  7.774e-06  &  -50.15642914  &  0.36567143 \\
12000   &  5.854e-06  &  -50.19615496  &  0.32594561 \\
			\hline
			$\infty$ &             &   -50.52210057  &            \\
			\hline
			\hline
		\end{tabular}
	\end{center}
\end{minipage}
\hfill
\begin{minipage}[t]{.48\textwidth}
	\begin{center}
		\begin{tabular}{wc{1.5cm}wc{1.8cm}wc{2cm}wc{3cm}}
			\multicolumn{4}{c}{} \\
			\multicolumn{4}{c}{$L=12$} \\
			\hline
			\hline
			$m$ &  $w_d$  & $E_0(m)$ &  $E_0(m)-E_0(\infty)$ \\
			\hline
 2000  &  -   &    -  &   - \\
 4000  &   1.840e-05  &  -71.40452905   &  1.73164686 \\
6000  &   1.486e-05  &  -71.71734724   &  1.41882867 \\
8000  &   1.365e-05  &  -71.91712273   &  1.21905318 \\
10000  &   1.224e-05  &  -72.06942431   &  1.06675160 \\
12000  &   9.293e-06  &  -72.19219444   &  0.94398147 \\
			\hline
			$\infty$ &             &   -73.13617591  &            \\
			\hline
			\hline
		\end{tabular}
	\end{center}
\end{minipage}
	\label{SBCdata}
\end{table}

We here check the accuracy of DMRG calculations under SBC (periodic chain).
In Table~\ref{SBCdata} the discarded weight $w_d$ and the ground-state energy
$E_0(m)$ of the half-filled square-lattice Hubbard model at $U/t=8$ are
shown as a function of the number of density-matrix eigenstates kept in
the renormalization $m$ for various system sizes $L \times L$. For all
cases the ground-state energy at $m\to\infty$ ($w_d\to0$) can be accurately
obtained by a linear fit of $E_0(m)$ vs. $w_d$. We also find that the
discarded weight is roughly proportional to $L$, i.e., $w_d \propto L$. Nevertheless,
the discarded weight is still order of $\sim10^{-5}$ even for $12 \times 12$.
This may fall into the category of an accurate DMRG calculation for 2D fermionic system.

In Table~\ref{DMRGdata} the discarded weight in the DMRG calculation for
the square-lattice Heisenberg model is compared between various boundary
conditions. The discarded weight is almost comparable for OBC and cylinder,
and somewhat larger for the SBC-projected periodic chain. Surprisingly,
it is much smaller when the SBC-projected open chain is studied. Thus,
from the viewpoint of discarded weight, OBC, SBC, and cylinder can be equally
good options in DMRG calculation for 2D systems

\newpage

\begin{table}[tbh]
	\caption{Discarded weight $w_d$ and the ground-state energy $E_0(m)$
		of the square-lattice Heisenberg model as a function of the number
		of density-matrix eigenstates kept in the renormalization $m$ for
		system sizes $6 \times 6$ ($L=6$) and $8 \times 8$ ($L=8$) with
		various boundary conditions. The ground-state energy at $m\to\infty$
		was obtained by a linear fit of $E_0(m)$ vs. $w_d$.}
\begin{minipage}[t]{.90\textwidth}
\begin{center}
\begin{tabular}{wc{1.8cm}|wc{1.8cm}wc{2cm}wc{2.5cm}|wc{1.8cm}wc{2cm}wc{2.5cm}}
				\multicolumn{7}{c}{OBC} \\
				\hline
				\hline
&\multicolumn{3}{c}{$L=6$}&\multicolumn{3}{c}{$L=8$} \\
$m$&$w_d$&$E_0(m)$&$E_0(m)-E_0(\infty)$&$w_d$&$E_0(m)$&$E_0(m)-E_0(\infty)$ \\
\hline
1000&1.100e-07&-21.72673267&0.00005340&8.166e-06&-39.60759545&0.01078493\\
2000&6.466e-09&-21.72678301&0.00000307&1.587e-06&-39.61627493&0.00210544\\
3000&9.883e-10&-21.72678556&0.00000051&5.714e-07&-39.61762743&0.00075294\\
4000&2.279e-10&-21.72678592&0.00000015&2.636e-07&-39.61803809&0.00034228\\
\hline
    &         &-21.72678607&          &         &-39.61838038&          \\
\hline
\hline
\end{tabular}
\end{center}
\end{minipage}

\vspace*{5.0mm}

\begin{minipage}[t]{.90\textwidth}
	\begin{center}
		\begin{tabular}{wc{1.8cm}|wc{1.8cm}wc{2cm}wc{2.5cm}|wc{1.8cm}wc{2cm}wc{2.5cm}}
			\multicolumn{7}{c}{PBC} \\
			\hline
			\hline
			&\multicolumn{3}{c}{$L=6$}&\multicolumn{3}{c}{$L=8$} \\
			$m$&$w_d$&$E_0(m)$&$E_0(m)-E_0(\infty)$&$w_d$&$E_0(m)$&$E_0(m)-E_0(\infty)$ \\
			\hline
1000&2.981e-05&-24.40302953&0.03665799&-&-&-\\
2000&9.188e-06&-24.42949472&0.01019281&-&-&-\\
3000&3.938e-06&-24.43541285&0.00427468&-&-&-\\
4000&1.982e-06&-24.43743236&0.00225516&-&-&-\\
			\hline
&         &-24.43968752&          & & &\\
			\hline
			\hline
		\end{tabular}
	\end{center}
\end{minipage}

\vspace*{5.0mm}

\begin{minipage}[t]{.90\textwidth}
	\begin{center}
		\begin{tabular}{wc{1.8cm}|wc{1.8cm}wc{2cm}wc{2.5cm}|wc{1.8cm}wc{2cm}wc{2.5cm}}
			\multicolumn{7}{c}{SBC (periodic chain)} \\
			\hline
			\hline
			&\multicolumn{3}{c}{$L=6$}&\multicolumn{3}{c}{$L=8$} \\
			$m$&$w_d$&$E_0(m)$&$E_0(m)-E_0(\infty)$&$w_d$&$E_0(m)$&$E_0(m)-E_0(\infty)$ \\
			\hline
1000&9.130e-06&-24.45300594&0.00801971&9.025e-05&-42.79118365&0.32899812\\
2000&1.661e-06&-24.45965335&0.00137230&4.565e-05&-42.96307806&0.15710371\\
3000&5.365e-07&-24.46058729&0.00043835&2.898e-05&-43.02157949&0.09860228\\
4000&2.260e-07&-24.46083548&0.00019017&1.973e-05&-43.05183014&0.06835163\\
			\hline
$\infty$&         &-24.46102565&          &         &-43.12018177&\\
			\hline
			\hline
		\end{tabular}
	\end{center}
\end{minipage}

\vspace*{5.0mm}

\begin{minipage}[t]{.90\textwidth}
	\begin{center}
		\begin{tabular}{wc{1.8cm}|wc{1.8cm}wc{2cm}wc{2.5cm}|wc{1.8cm}wc{2cm}wc{2.5cm}}
			\multicolumn{7}{c}{SBC (open chain)} \\
			\hline
			\hline
			&\multicolumn{3}{c}{$L=6$}&\multicolumn{3}{c}{$L=8$} \\
			$m$&$w_d$&$E_0(m)$&$E_0(m)-E_0(\infty)$&$w_d$&$E_0(m)$&$E_0(m)-E_0(\infty)$ \\
			\hline
1000&1.037e-09&-23.07464567&0.00000050&9.819e-07&-41.31375095&0.00116164\\
2000&2.447e-11&-23.07464616&0.00000001&1.161e-07&-41.31476448&0.00014811\\
3000&2.351e-12&-23.07464617&$<10^{-8}$&2.987e-08&-41.31487486&0.00003773\\
4000&3.837e-13&-23.07464618&$<10^{-8}$&1.058e-08&-41.31489881&0.00001379\\
			\hline
			$\infty$&         &-23.07464618&          &         &-41.31491259&\\
			\hline
			\hline
		\end{tabular}
	\end{center}
\end{minipage}

\vspace*{5.0mm}

\begin{minipage}[t]{.90\textwidth}
	\begin{center}
		\begin{tabular}{wc{1.8cm}|wc{1.8cm}wc{2cm}wc{2.5cm}|wc{1.8cm}wc{2cm}wc{2.5cm}}
			\multicolumn{7}{c}{cylinder} \\
			\hline
			\hline
			&\multicolumn{3}{c}{$L=6$}&\multicolumn{3}{c}{$L=8$} \\
			$m$&$w_d$&$E_0(m)$&$E_0(m)-E_0(\infty)$&$w_d$&$E_0(m)$&$E_0(m)-E_0(\infty)$ \\
			\hline
1000&3.349e-08&-23.09009340&0.00001678&4.299e-06&-41.34358356&0.00569929\\
2000&1.512e-09&-23.09010942&0.00000076&7.206e-07&-41.34831869&0.00096415\\
3000&2.012e-10&-23.09011008&0.00000010&2.330e-07&-41.34898086&0.00030198\\
4000&4.556e-11&-23.09011016&0.00000002&9.196e-08&-41.34916154&0.00012131\\
			\hline
			$\infty$&         &-23.09011018&          &         &-41.34928284&\\
			\hline
			\hline
		\end{tabular}
	\end{center}
\end{minipage}
	\label{DMRGdata}
\end{table}

\newpage

\section{Advantages of spiral boundary conditions}

\begin{figure}[h]
	\includegraphics[width=0.9\linewidth]{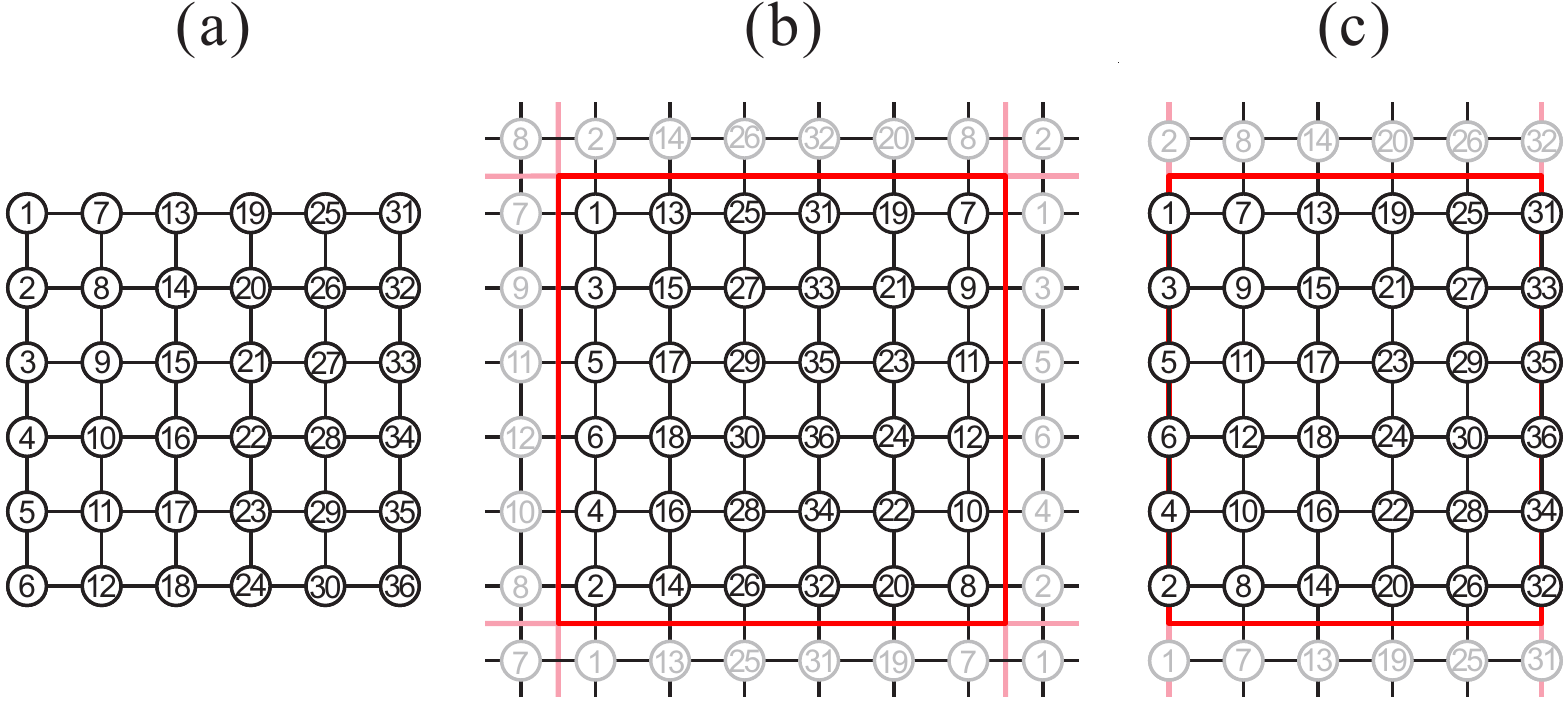}
	\caption{2D square-lattice $6\times6$ cluster with (a) OBC, (b) PBC,
		and (c) cylinder. In each cluster an optimal order of sites for
		DMRG simulation is denoted by numbers.
	}
	\label{fig:lattice6x6_SM}
\end{figure}

In DMRG simulations we have to choose a proper boundary condition for
considered system and quantities. The applicability of DMRG to a lattice
system can be roughly judged by the distance between two sites connected
by the longest-range term in the system. The distance is denoted by $d$ below.
In general, the DMRG calculation is difficult if $d$ is larger than $\sim10$.
Let us then estimate the values of $d$ for various boundary conditions. In
Fig.~\ref{fig:lattice6x6_SM} an optimal order of sites to give the shortest
$d$ for a 2D square-lattice cluster is illustrated for each boundary condition. 
When we use OBC or cylinder, $d$ is $L$ for a $L \times L$ cluster. This is comparable to that for SBC ($d=L-1$ or $d=L$). However, if we use PBC, $d$
jumps to $2L$. This is definitely ill-conditioned for DMRG treatments.
For example, the system size is limited up to $L\sim6$ in the Heisenberg model.
Therefore, PBC may be firstly excluded from the choice, although it is
naively thought to be a most versatile boundary condition in numerical
calculations.

In fact, either OBC or cylinder is typically employed in the previous DMRG
simulations. These boundary conditions are technically rather convenient
for DMRG simulations because the value of $d$ can be reduced down to $L$
with the optimal order of sites. Nevertheless, they could often lead to some
sorts of problems in the physical properties. The followings are examples of
issues: With OBC, the bulk state can be significantly disturbed by Friedel
oscillations, and also the properties of bulk and edge states can be completely
different due to the missing outside bonds especially in doped systems.
These problems could be managed by controlling open edges for a specific
states, and however, it is not always successful. With cylinder, 
a kind of spatial anisotropy is introduced in spite of the original isotropic
2D square-lattice cluster. A most problematic point is that short loops of
bonds are formed along the circumference direction. This could results in
an unnatural periodicity of the wave function as well as an unexpected
plaquette constraint of particles or spins. Accordingly, as shown in the
previous section of Supplemental Material, the density-density correlation
functions can be largely disturbed by cylindrical boundary.

With SBC, Friedel oscillations can be avoidable and no artificial short loops
of bonds are created. Also, $d$ if only $L$ or $L-1$. Furthermore,
a SBC-projected chain with open boundary is used, the accuracy of DMRG
calculation is even better than that with OBC and cylinder as shown in the
previous section of Supplemental Material. Thus, SBC enables us to perform
accurate DMRG calculations with skirting some sorts of finite-size effects
seen in OBC and cylindrical clusters. However, we expect that SBC does not
always provide better results. For example, OBC may be more appropriate to
study a system exhibiting valence-bond-solid state such as plaquette/bond
order; and, a cylindrical cluster may be more appropriate to study a system
exhibiting magnetic/charge/bond ordered state with spatial rotation symmetry
breaking.

We thus think that SBC can be a main option of the boundary conditions in
DMRG calculations for the above technical and physical reasons.